\documentclass[iop]{emulateapj}
\usepackage{amssymb,amsmath,amsfonts,amsbsy}
\usepackage{color}
\usepackage{natbib}
\usepackage{enumerate}
\usepackage{graphicx}
\usepackage{bbm}
\newcommand\id{\ensuremath{\mathbbm{1}}} 
\def\nn{\nonumber}

\def\dA{{\rm d}A}

\def\dV{{\rm d}V}
\def\du{{\rm d}u}

\begin{document}
\title{Minkowski Tensors in Three Dimensions - Probing the Anisotropy Generated by Redshift Space Distortion}

\author{Stephen Appleby$^{a}$}\email{stephen@kias.re.kr}
\author{Pravabati Chingangbam$^{b}$}
\author{Changbom Park$^{a}$}
\author{K. P. Yogendran$^{c,d}$}
\author{Joby P. K.$^{b,e}$}
\affiliation{$^{a}$School of Physics, Korea Institute for Advanced Study, 85 Hoegiro, Dongdaemun-gu, Seoul 02455, Korea}
\affiliation{$^{b}$Indian Institute of Astrophysics, Koramangala II Block, Bangalore 560 034, India}
\affiliation{$^{c}$Indian Institute of Science Education and Research, Sector 81, Mohali, India}
\affiliation{$^{d}$Indian Institute of Science Education and Research, C/o Sree Rama Engineering College (Transit Campus), Karakambadi Road, Mangalam (P.O.), Tirupati 517 507, India}
\affiliation{$^{e}$Department of Physics, University of Calicut, Malappuram, Kerala-673 635, India}

\begin{abstract}
We apply the Minkowski tensor statistics to three dimensional Gaussian random fields. Minkowski tensors contain information regarding the orientation and shape of excursion sets, that is not present in the scalar Minkowski functionals. They can be used to quantify globally preferred directions, and additionally provide information on the mean shape of subsets of a field. This makes them ideal statistics to measure the anisotropic signal generated by redshift space distortion in the low redshift matter density field. We review the definition of the Minkowski tensor statistics in three dimensions, focusing on two coordinate invariant quantities $W^{0,2}_{1}$ and $W^{0,2}_{2}$. We calculate the ensemble average of these $3 \times 3$ matrices for an isotropic Gaussian random field, finding that they are proportional to products of the identity matrix and a corresponding scalar Minkowski functional. We show how to numerically reconstruct $W^{0,2}_{1}$ and $W^{0,2}_{2}$ from discretely sampled fields and apply our algorithm to isotropic Gaussian fields generated from a linear $\Lambda$CDM matter power spectrum. We then introduce anisotropy by applying a linear redshift space distortion operator to the matter density field, and find that both $W^{0,2}_{1}$ and $W^{0,2}_{2}$ exhibit a distinct signal characterised by inequality between their diagonal components. We discuss the physical origin of this signal and how it can be used to constrain the redshift space distortion parameter $\Upsilon \equiv f/b$. 
\end{abstract}

\maketitle

\section{\label{sec:1}Introduction}

The Minkowski tensors are a set of statistics which generalise the Minkowski functionals, which are themselves scalar quantities. They are defined as integrals over the boundary of an excursion set, with integrands related to symmetric tensor products of position vectors and normals to the boundary \citep{had71,schneid72,McMullen:1997,Alesker1999,2002LNP...600..238B,HugSchSch07,1367-2630-15-8-083028,JMI:JMI3331}. As such, they provide directional information that is not present in the scalar Minkowski functionals. 

The Minkowski functionals have been used within the context of cosmology for three decades \citep{1970Ap......6..320D,Adler,Gott:1986uz,Hamilton:1986, Gott:1986uz,Ryden:1988rk,1989ApJ...340..625G,1989ApJ...345..618M,1992ApJ...387....1P,1991ApJ...378..457P, Matsubara:1994we,1996ApJ...457...13M,Schmalzing:1995qn,Gott:2006yy,Gott:2008kk,0004-637X-529-2-795,2005ApJ...633....1P,Kerscher:2001gm, Appleby:2017ahh, Appleby:2018jew}. However, the application of Minkowski tensors to the field is a relatively new phenomenon. In a recent publication, the authors analytically calculated the ensemble expectation value of the Minkowski tensor statistic $W^{1,1}_{2}$ for two dimensional Gaussian random fields on a sphere or plane \citep{Chingangbam:2017uqv}. This quantity is a $2\times 2$ matrix, and for an isotropic field is proportional to the identity matrix. In an arbitrary coordinate system, both an anisotropic Gaussian field and isotropic non-Gaussian field can potentially generate off-diagonal elements. By diagonalising these matrices, we eliminate the coordinate dependence and find that an isotropic, non-Gaussian field would yield equal eigenvalues that differ from their Gaussian expectation values. An anisotropic field would yield unequal eigenvalues.

These statistics were first applied to cosmological data in \citet{2017JCAP...06..023G}, where a significant anisotropic signal was found in the 2015 $E$-mode Planck data \citep{Adam:2015tpy}. The authors then applied this statistic to non-Gaussian density fields constructed from slices of mock galaxy simulations in \citet{Appleby:2017uvb}. The density field reconstructed from mock galaxies contains a preferred direction as a result of redshift space distortion - this effect produced a distinct imprint on the diagonal elements of $W^{1,1}_{2}$. These statistics have also been applied to the fields of the epoch of reionization \citep{Kapahtia:2017qrg}.

In \citet{Appleby:2017uvb} the three-dimensional matter density field was decomposed into two dimensional slices. This approach is useful when dealing with photometric redshift catalogs, which are characterised by large galaxy number density and volume, but relatively poor position information along the line of sight. However, with increasingly large spectroscopic galaxy catalogs becoming available \citep{doi:10.1093/mnras/stv1436, 2016AJ....151...44D,2015ApJ...798....7B, 2017AJ....154...28B}, we can directly extract information from the full three dimensional dark matter density field. The three dimensional field contains more information than its two dimensional counterpart - the process of binning galaxies into two dimensional slices smooths the distribution and information is lost in the Fourier modes parallel to the line of sight. In addition, we expect the two and three dimensional statistics to exhibit different sensitivity to bias, shot noise and redshift space distortion. The effect of linear redshift space distortion is to modify the shape of large scale structures in the direction parallel to the line of sight. The Finger of God effect will scatter the line of sight position of galaxies within virialised structures as a result of stochastic velocity components. Both effects will change the morphology of structures, generating anisotropic iso-density contours. The strength of the anisotropic signal can be used to test the nature of gravity and additionally constrain cosmological parameters. Many previous studies on redshift space distortion have focused on how the signal modifies the shape of the matter power spectrum (and also potentially the bispectrum). However, the Minkowski tensors provide a measure of the shape of excursion sets, and hence can be used to probe the anisotropic effect of the velocity field on structures. 

In this work we extend our previous analytic and numerical analysis of two dimensional Minkowski tensors to three dimensional fields. We begin in section \ref{sec:2} by reviewing the definition of Minkowski functionals and their generalisation to tensors. We then focus on two particular Minkowski tensors - $W^{0,2}_{1}$, $W^{0,2}_{2}$ - that possess translational invariance, and calculate their ensemble average for a Gaussian random field.  

To extract the $W^{0,2}_{1}$, $W^{0,2}_{2}$ statistics from a discretely sampled density field we require a method of numerically generating iso-density surfaces. The technical details of the algorithm adopted in this work are discussed in appendices \ref{sec:3}, \ref{sec:error}.  We apply this numerical algorithm to a Gaussian random field in section \ref{sec:4}, matching our numerical analysis with analytic predictions where possible. We study the shape of individual connected regions and holes in section \ref{sec:5} by calculating the eigenvalues of $W^{0,2}_{1}$, $W^{0,2}_{2}$ for disjoint excursion sub-sets. In section \ref{sec:rsd} we repeat our analysis for an anisotropic Gaussian field, by applying a linear redshift space distortion operator to the matter density field. We discuss the sensitivity of $W^{0,2}_{1}$, $W^{0,2}_{2}$ to the redshift space distortion parameter $\Upsilon = f/b$ and summarize our findings in section \ref{sec:sum}.

\section{Minkowski Tensors} 
\label{sec:2}

In this section we review the definition and properties of Minkowski functionals in three dimensions, and their generalisation to Minkowski tensors. 

The Minkowski functionals are a set of $d+1$ scalar quantities that characterize the morphology and topology of an excursion set of a field $u$ in $d$-dimensional space. Throughout this work we use mean subtracted fields $\langle u \rangle =0$ with variance $\sigma_{0}^{2} =\langle u^{2} \rangle$. An excursion set is defined by a boundary of constant field value $u=\sigma_{0}\nu={\rm constant}$. In three dimensions the excursion set boundary is a two-dimensional iso-field surface, and the Minkowski functionals are the enclosed volume and surface area, and integrals of the mean and Gaussian curvatures over the surface. Specifically we can write 

\begin{eqnarray} \label{eq:a1} & & W_{0} = {1 \over V} \int_{Q} dV  \\ 
\label{eq:a2} & & W_{1} = {1 \over 6 V} \int_{\partial Q} dA  \\
\label{eq:a3} & & W_{2} = {1 \over 3\pi V} \int_{\partial Q} G_{2} dA  \\ 
\label{eq:a4} & & W_{3} = {1 \over 4\pi^{2} V} \int_{\partial Q} G_{3} dA   \end{eqnarray}

\noindent where $Q$ is the excursion set volume and $\partial Q$ its boundary. $dV$, $dA$ are infinitesimal volume and surface area elements respectively, and $G_{2} = (\kappa_{1}+\kappa_{2})/2$, $G_{3} = \kappa_{1} . \kappa_{2}$ are the mean and Gaussian curvatures of the surface $\partial Q$ respectively. $\kappa_{1}, \kappa_{2}$ are the principle curvatures of the surface and $V$ is the total volume of the three dimensional space; we have defined the Minkowski functionals per unit volume. 

For a Gaussian random field, one can analytically calculate the ensemble expectation value of these quantities \citep{doi:10.1143/PTP.76.952,1970Ap......6..320D, Adler, Gott:1986uz, Hamilton:1986} -

\begin{eqnarray} \label{eq:a5} & & \langle W_{0} \rangle = {1 \over 2} {\rm erfc} \left({\nu \over \sqrt{2}}\right)  \\ 
\label{eq:a6} & & \langle W_{1} \rangle = {1 \over 3\sqrt{3}\pi}{\sigma_{1} \over \sigma_{0}}e^{-\nu^{2}/2}  \\
\label{eq:a7} & & \langle W_{2} \rangle = {1 \over 9 \sqrt{2}\pi^{3/2}} {\sigma_{1}^{2} \over  \sigma_{0}^{2}}\nu e^{-\nu^{2}/2}  \\ 
\label{eq:a8} & & \langle W_{3} \rangle = {1 \over 4\pi^{2}}\left({\sigma_{1} \over \sqrt{3}\sigma_{0}}\right)^{3} e^{-\nu^{2}/2}\left(1-\nu^{2}\right)   \end{eqnarray}

\noindent where $\sigma_{0,1}$ are the two-point cumulants of the field - 

\begin{equation} \label{eq:cumulant} \sigma^{2}_{i} = \int {k^{2} dk \over 2\pi^{2}} k^{2i} P(k) \end{equation}

\noindent and $P(k)$ is the power spectrum from which $u(x,y,z)$ is drawn. The $\nu$ dependence of $W_{0-3}$ is completely fixed, and the amplitudes carry information regarding the shape of the power spectrum. For a non-Gaussian field - such as the late Universe matter density which has undergone non-linear gravitational collapse - both the amplitude and shape of these functions is modified \citep{Matsubara:1994we, 2000astro.ph..6269M,Pogosyan:2009rg, Gay:2011wz,Codis:2013exa}. Specifically $W_{0-3}$ lose their symmetry properties about $\nu=0$ due to the presence of corrections proportional to higher order cumulants of the field. 

These statistics can be generalised to vectors and tensors \citep{had71,schneid72,McMullen:1997,Alesker1999,2002LNP...600..238B,HugSchSch07,1367-2630-15-8-083028,JMI:JMI3331}. Our focus is on rank-two statistics, although rank one and higher rank quantities also carry information. The rank-two Minkowski tensors in three dimensions are defined as \citep{1367-2630-15-8-083028}

\begin{eqnarray}\label{eq:mtd1} & & W^{2,0}_{0} \equiv A_{0} \int_{Q} {\bf x}^{2} dV \\ 
\label{eq:mtd2} & & W^{r, s}_{t} \equiv A_{t} \int_{\partial Q} G_{t} {\bf x}^{r} \otimes {\bf \hat{n}}^{s} dA \end{eqnarray}

\noindent where $t = 1,2,3$ and $(r,s)=(2,0),(1,1)$ or $(0,2)$. ${\bf x}$ and ${\bf \hat{n}}$ are the position vector and unit normal to the bounding surface, respectively, and $G_{1}=1$, $G_{2} = (\kappa_{1} + \kappa_{2})/2$ and $G_{3} = \kappa_{1} . \kappa_{2}$. ${\bf x}^{r}$, ${\bf \hat{n}}^{s}$ are the symmetric tensor products of ${\bf x}$, ${\bf \hat{n}}$ with themselves $r$, $s$ times. $A_{t}$ is a normalisation factor chosen to match the scalar Minkowski functionals, so $A_{0}=1/V$ $A_{1} = 1/6V$, $A_{2}=1/3\pi V$ and $A_{3} = 1/4\pi^{2}V$. The definitions ($\ref{eq:mtd1},\ref{eq:mtd2}$) comprise a complete set of $3\times 3$ matrices that can be constructed from ${\bf x}$ and ${\bf \hat{n}}$ integrated over $Q$ and $\partial Q$. All other rank two-quantities that can be constructed vanish identically. 

There are ten Minkowski tensors defined in this way, however they are not all independent. The following four linear dependencies exist \cite{HUG2008482,HugSchSch07} 

\begin{equation} W_{t} \id = t W^{0,2}_{t} + {A_{t+1} \over A_{t}}(3-t) W^{1,1}_{t+1} \end{equation}

\noindent for $t=0,1,2,3$, where $W_{t}$ are the scalar Minkowski functionals and $\id$ is the $3 \times 3$ identity matrix. We can therefore form a basis of six Minkowski tensors which encapsulate all relevant shape information that can be extracted from the rank-2 matrices - $W^{2,0}_{0}, W^{2,0}_{1}, W^{2,0}_{2}, W^{2,0}_{3}, W^{0,2}_{1}, W^{0,2}_{2}$. Of these quantities, only $W^{0,2}_{1}$ and $W^{0,2}_{2}$ are translationally invariant \citep{1367-2630-15-8-083028}. The remaining four will vary as a function of the coordinate system that one adopts. For cosmological applications a coordinate system centered on the observer presents a logical choice, but in what follows we will focus on the statistics $W^{0,2}_{1}$ and $W^{0,2}_{2}$ with the understanding that we are not extracting all possible information from the field. 

The Minkowski tensors that we study, which are explicitly given by 

\begin{eqnarray} \label{eq:w021} & & W^{0,2}_{1} = {1 \over 6 V} \int_{\partial Q} {\bf \hat{n}}^{2} dA , \\ 
 \label{eq:w022} & & W^{0,2}_{2} = {1 \over 3\pi V} \int_{\partial Q} G_{2} {\bf \hat{n}}^{2} dA , \end{eqnarray} 

\noindent represent integrals over the boundary of the excursion set, with the vector product of surface normal ${\bf \hat{n}}$ with itself as the kernel. They are $3\times 3$ matrices, and as we will show both the structure of the matrix and the magnitude of its components inform us about the properties of the field.

\begin{figure*}[!b]
  \includegraphics[width=1.0\textwidth]{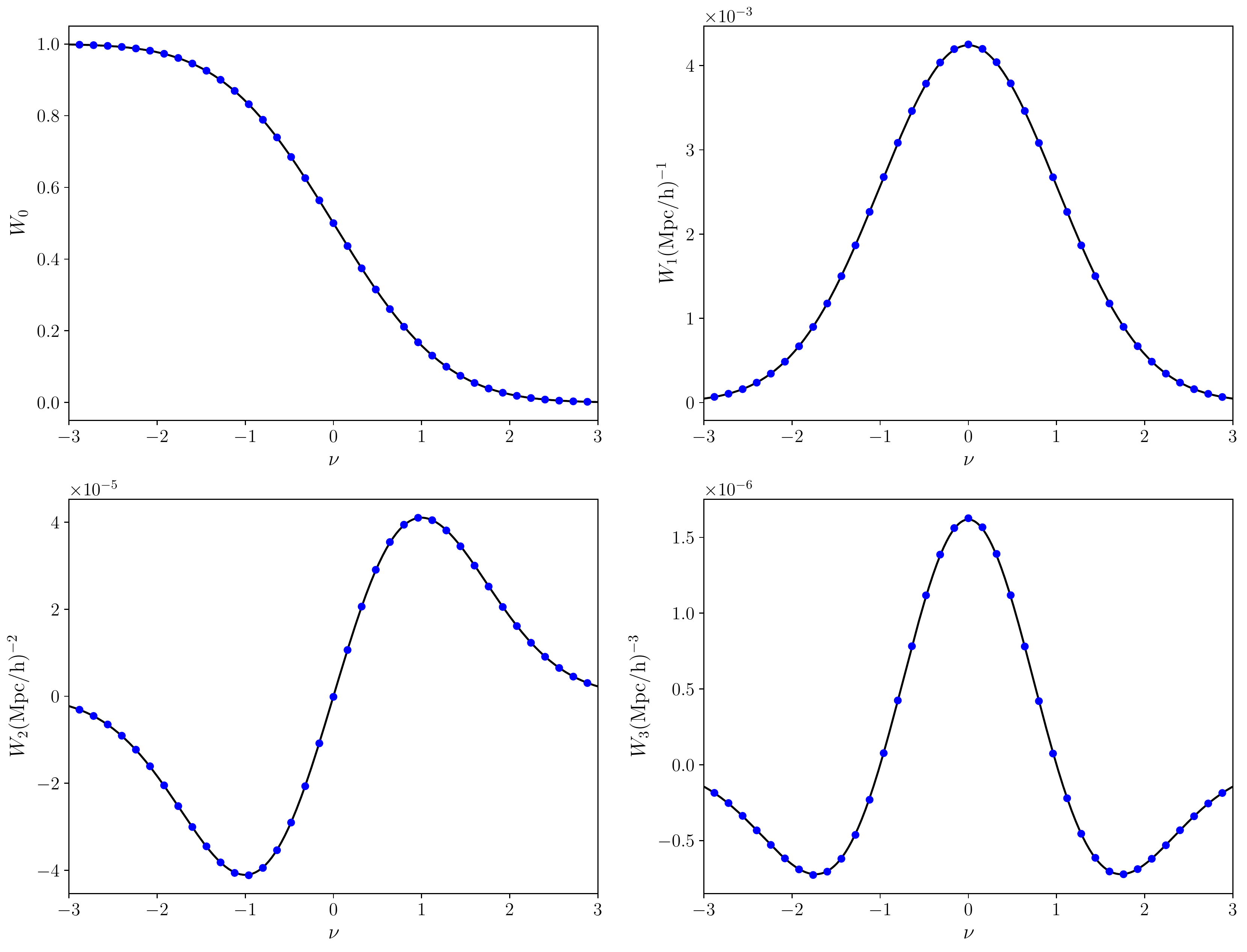}
  \caption{The Minkowski functionals. We generate $N_{\rm real } = 50$ realisations of a Gaussian field drawn from a $\Lambda$CDM power spectrum in a cube of volume $V = \left( 1024 h^{-1} \, {\rm Mpc} \right)^{3}$ and resolution $\epsilon = 2 h^{-1} \, {\rm Mpc}$. We smooth the field with a Gaussian kernel with smoothing scale $R_{\rm G} = 20 h^{-1} \, {\rm Mpc}$. The mean and error on the mean are exhibited as blue points and error bars respectively -- the error bars are smaller than the data points. The black solid line denotes the theoretical expectation value for the field, which shows close agreement to the numerical reconstruction. }
  \label{fig:8}
\end{figure*}

For a Gaussian random field, we can calculate the ensemble expectation value of these matrices. We close this section with an outline of this calculation.

We define $u$ as a smooth random field in three-dimensional space and $\nu$ is a threshold value which defines the excursion set boundary $\partial Q$. We first re-write the integrands of $W_1^{0,1}$ and $W_2^{0,2}$ in terms of the field $u$ and its first and second derivatives, $u_i$ and $u_{ij}$. $i, j$ subscripts run over a Cartesian $(x_{1},x_{2},x_{3})$ coordinate system. The unit normal vector is given by $\mathbf{\hat n}= \nabla u/|\nabla u|$.  For an arbitrary integrand $F(\nu)$, we can transform area integrals to volume integrals by introducing an integral over $u$ with a delta function, $\delta(u-\nu)$, as,
\begin{equation}
  \int \dA\,F(\nu) =   \int \dA\,\du \, \delta(u-\nu)\,F(u) =  \int \dV \,|\nabla u|\, \delta(u-\nu)\, F(u).
\end{equation}

The mean curvature is related to the unit vector normal to a given surface as,

    \begin{equation}
    G_2 = -\frac{1}{2} \, \nabla \cdot \mathbf{\hat{n}}.
    \end{equation}

In terms of field derivatives, we can express $G_2$ as

\begin{widetext}
\begin{equation}
  G_2 = \frac{1}{2\left(u_1^2+u_2^2+u_3^2\right)^{3/2}}\bigg[ 2\bigg(u_1 u_2 u_{12} +
   u_2 u_3 u_{23}+ u_1 u_3 u_{13}\bigg) -\bigg(u_1^2(u_{22}+u_{33}) + u_2^2(u_{11}+u_{33}) + u_3^2(u_{11}+u_{22})\bigg) \bigg] .
\end{equation}
\end{widetext}

$W_1^{0,2}$ and  $W_2^{0,2}$  are given by 

\begin{eqnarray}
   W_1^{0,2}  &=& A_1\,\int  \dV  \ \delta(u-\nu)\ \frac{1}{|\nabla u|} \ {\mathcal M},\\
    W_2^{0,2} &=& A_2\,\int  \dV \ \delta(u-\nu)\ \frac{G_2}{|\nabla u|} \ {\mathcal M},
\end{eqnarray}
where the matrix $\mathcal M$ is given by
\begin{equation}
  \mathcal M=  \left(
  \begin{array}{ccc} 
    u_1^2 &  u_1 \,u_2 &  u_1 \,u_3 \\
    u_1 \,u_2 & u_2^2 &  u_2 \,u_3 \\
        u_1 \,u_3 & u_2 u_3  & u_3^2
  \end{array}\right).
  \label{eqn:M}
  \end{equation}

The trace of $W_1^{0,2}$ is equal to $W_1$ -
\begin{eqnarray}
  \sum_i \left( W_1^{0,2}\right)_{ii} &=& A_1\,\int  \,\dV\, \delta(u-\nu)\
  \frac{1}{|\nabla u|} \ \left( u_1^2 + u_2^2 + u_3^2\right) \nn\\ 
       {}& =& A_1\,\int  \,\dV\, \delta(u-\nu)\ |\nabla u|\nn\\
       {} &=& W_1 ,     
\end{eqnarray}

\noindent and the trace of $W_2^{0,2}$ equivalent to $W_2$ - 
\begin{eqnarray}
  \sum_i \left( W_2^{0,2}\right)_{ii} &&= A_2\,\int \,\dV\, \delta(u-\nu)\
  \frac{G_2}{|\nabla u|} \ \left( u_1^2 + u_2^2 + u_3^2\right) \nn\\
       {}&=& A_2\,\int \,\dV\, \delta(u-\nu)\
  {G_2}\,{|\nabla u|} \nn\\
  {} &=& W_2.
\end{eqnarray}

\vskip 1.5cm
\subsection{Ensemble expectation values for isotropic Gaussian fields}
\label{sec:gaussian}

If $u$ is Gaussian and isotropic we can compute the ensemble expectation values of $W_j^{0,2}$. $\sigma_{0,1}$ are defined in terms of the variances of $u$ and $u_{i}$ as, 
\begin{eqnarray}
  \sigma_0^2 &\equiv& \langle u^2\rangle ,\\
  \sigma_1^2 &\equiv& \langle |\nabla u|^2\rangle ,
\end{eqnarray}

\noindent which are the real space equivalents of the first two cumulants defined in equation ($\ref{eq:cumulant}$). The non-zero correlations containing second derivatives are
\begin{eqnarray}
  \nonumber & &   \langle u_{ii} u_{ii}\rangle \equiv c, \quad 
  \langle u_{ij} u_{ij}\rangle_{i\ne j} \equiv s , \\
 & &   \langle u_{11} u_{22}\rangle = \langle u_{22} u_{33}\rangle = \langle u_{11} u_{33}\rangle \equiv q.  
  \end{eqnarray}

  The ten component column vector
$\vec{X}=(u,\,u_1,\,u_2,\,u_3,\,u_{11},\,u_{22},\,u_{33},\,u_{12}, \,u_{23}, \,u_{13})$
 has a joint probability distribution of the form
\begin{equation}
  P(\vec{X}) = \frac1{\sqrt{(2\pi)^{10} \ \rm{\mathbf{Det}} \Sigma }} \,\exp\left(-\frac12 \,\vec{X}^{\rm{\mathbf{T}}}\,{\Sigma}^{-1}\,\vec{X} \right), 
\label{eqn:gpdf}
\end{equation}
where the covariance matrix $\Sigma \equiv \langle \vec{X}\vec{X}\,{}^{\mathbf T} \rangle$ is given by
\begin{widetext}
\begin{equation}
  \Sigma = \left(
  \begin{array}{cccccccccc}
    \sigma_0^2 & 0 & 0 & 0 & - \frac{\sigma_1^2}{3} & -\frac{\sigma_1^2}{3} & -\frac{\sigma_1^2}{3} & 0 & 0 & 0\\
    0 & \frac{\sigma_1^2}{3} & 0 & 0 & 0 & 0 & 0 & 0 & 0 & 0 \\
    0 & 0 & \frac{\sigma_1^2}{3} & 0 & 0 & 0 & 0 & 0 & 0 & 0 \\
    0 & 0 & 0 & \frac{\sigma_1^2}{3} & 0 & 0 & 0 & 0 & 0 & 0 \\
    -\frac{\sigma_1^2}{3} & 0 & 0 & 0 & c & q & q & 0 & 0 & 0\\
    -\frac{\sigma_1^2}{3} & 0 & 0 & 0& q & c & q & 0 & 0 & 0\\
    -\frac{\sigma_1^2}{3}  & 0 & 0 & 0 & q & q &  c  &  0 & 0 & 0 \\
    0 & 0 & 0 & 0 & 0 & 0 & 0 & s  & 0 & 0\\
    0 & 0 & 0 & 0 & 0 & 0 & 0 & 0 & s &0\\
    0 & 0 & 0 & 0 & 0 & 0  & 0 & 0& 0& s 
  \end{array}
  \right)
\end{equation}
\end{widetext}
The ensemble expectation values of $W_1^{0,2}$ and $W_2^{0,2}$ are

\begin{eqnarray}
 \nonumber  \big\langle W_1^{0,2}(\nu) \big\rangle
               &=& A_1\int \dV \,\int {\rm d}\vec{X} \,P(\vec{X})\,\delta(u-\nu) \,\frac{1}{|\nabla u|}  \,{\mathcal M},\\
    & & \\
  \nonumber   \big\langle W_2^{0,2}(\nu) \big\rangle &=& A_2\int\dV\,\int {\rm d}\vec{X} \,P(\vec{X})\,\delta(u-\nu) \,\frac{G_2}{|\nabla u|}  \,{\mathcal M}. \\ 
   & & 
    \label{eqn:ensemblex}
\end{eqnarray}

\noindent The integrals over volume and $\vec X$ have been interchanged since they commute. Performing the $\vec X$ integration first, we arrive at ensemble expectation per unit volume

\begin{eqnarray}
  \big\langle  W_1^{0,2}(\nu) \big\rangle&=&
  \frac{\sigma_{1}}{9\sqrt{3}\,\pi\ \sigma_{0}}\, e^{-\nu^2/2} \ \id 
=  \frac{ \langle W_{1} \rangle }{3} \ \id 
  \\
   \big\langle  W_2^{0,2}(\nu) \big\rangle&=&
    \frac{\sigma_{1}^{2}}{27\pi \sqrt{2\pi} \ \sigma_{0}^2} \ \nu \,e^{-\nu^2/2} \ \id
   = \frac{ \langle W_{2} \rangle }{3} \ \id
   \end{eqnarray}

When extracting information from the Minkowski tensors, one can either use the components of the matrices $W^{0,2}_{1}$ and $W^{0,2}_{2}$ directly, or alternatively construct their eigenvalues and eigenvectors. Any anisotropic signal in the field will generate inequality between eigenvalues of these matrices, independent of the particular coordinate system adopted.

\begin{table}
\begin{center}
\caption{\label{tab:ii}  }
 \begin{tabular}{||c | c ||}
 \hline
 Parameter & Fiducial Value \\ [0.5ex] 
 \hline\hline
 $\Omega_{\rm mat}$ & $0.26$   \\ 
 \hline
 $\Omega_{\Lambda}$ & $0.74$   \\ 
 \hline
 $H_{0}$ & $72 {\rm km /s/Mpc}$ \\
 \hline
 $R_{\rm G}$ & $20\, h^{-1} {\rm Mpc}$   \\
 \hline
 $\epsilon$ & $2\, h^{-1} {\rm Mpc}$   \\
 \hline
\end{tabular}
\end{center} 
Fiducial parameters used in this work. $R_{\rm G}$ is the smoothing scale used when applying a Gaussian smoothing kernel to the density field. $\epsilon$ is the fiducial resolution of the density fields generated in this work. 
\end{table}

\begin{figure*}[!b]
  \includegraphics[width=1.0\textwidth]{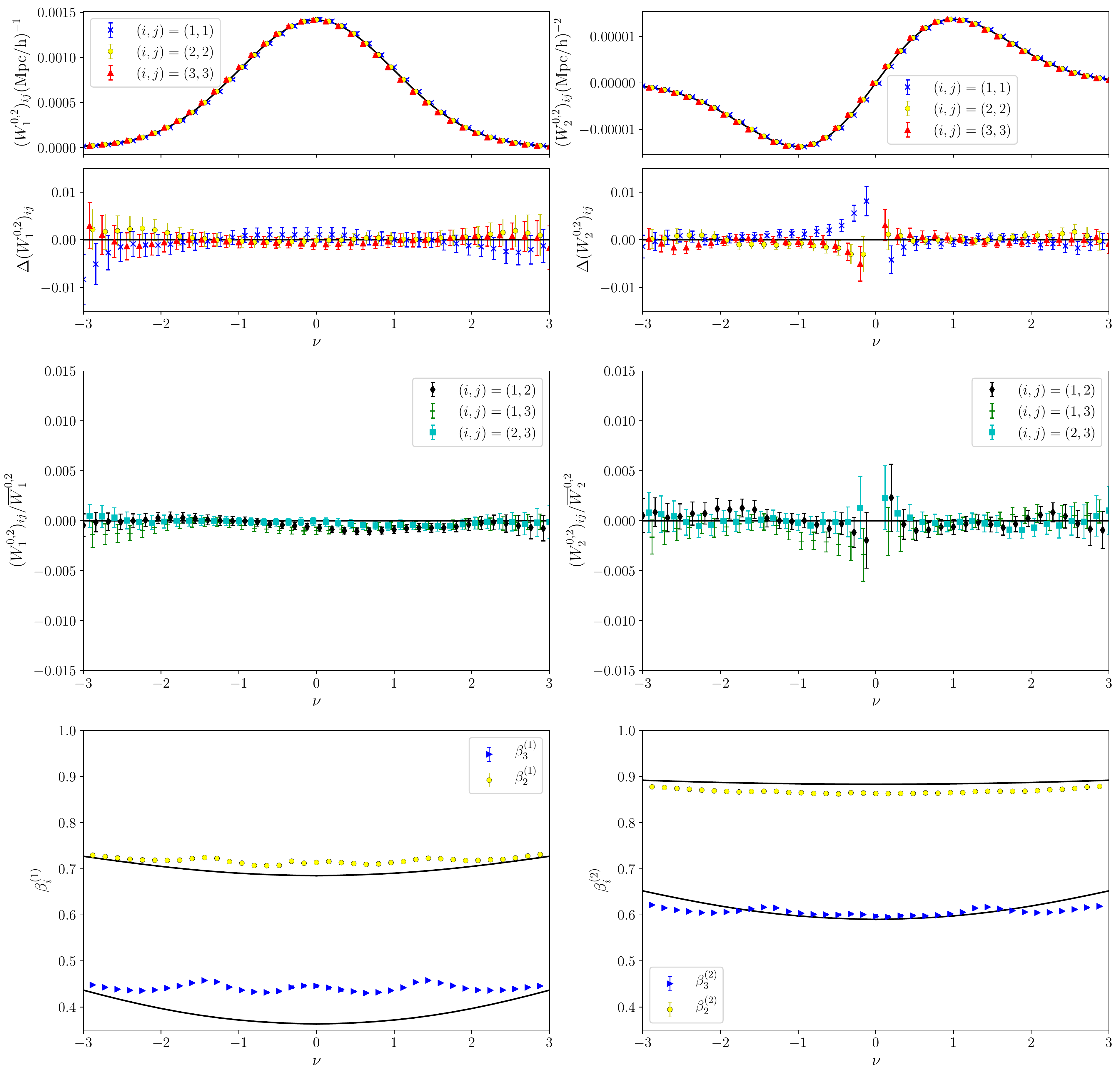} 
  \caption{The Minkowski tensors $W^{0,2}_{1}$ (left panels), $W^{0,2}_{2}$ (right panels). [Top panels] diagonal components of the $W^{0,2}_{1}$, $W^{0,2}_{2}$ matrices $(i,j)=(1,1),(2,2),(3,3)$ (blue/yellow/red) and the theoretical expectation value (black solid line). We also plot the residuals $\Delta (W^{0,2}_{1})_{ij}$, $\Delta (W^{0,2}_{2})_{ij}$ defined in equations ($\ref{eq:dw1},\ref{eq:dw2}$), which are consistent with zero. [Middle panels] the ratio of the off-diagonal components of $W^{0,2}_{1}$, $W^{0,2}_{2}$ and the average of the diagonal components. These quantities are consistent with zero as expected for a Gaussian field. [Bottom panels] $\beta^{(1)}_{i}$ and $\beta^{(2)}_{i}$, where $i=2,3$. The yellow,blue points correspond to $i=2$,$i=3$ respectively. The $\beta^{(1,2)}_{i}$ functions are defined in equations ($\ref{eq:bet1},\ref{eq:bet2}$) - they provide a measure of the mean shape of connected regions and holes in the excursion set. The solid black lines in the figures correspond to the $\beta^{(1,2)}_{i}$ values predicted using the shape of peaks in a Gaussian random field.}
  \label{fig:MT}
\end{figure*}

\section{Numerical Calculation of Minkowski Tensors for Isotropic Gaussian Random Fields}
\label{sec:4}

We have argued that additional information is potentially contained in the Minkowski tensors relative to the Minkowski functionals, and calculated their theoretical expectation value for an isotropic Gaussian random field. We now numerically extract these statistics from a discretely sampled density field. To do so, we require an algorithm capable of generating bounding surfaces of constant density enclosing an excursion set. In appendix \ref{sec:3} we detail our method of surface generation, and how we calculate the Minkowski tensors using the surface. 

In this work we study three dimensional density fields $\delta_{i,j,k}$ on a regular lattice, where $i,j,k$ are integer pixel identifiers in a $x_{1},x_{2},x_{3}$ Cartesian coordinate system. We take $1 \le i,j,k \le N_{\rm pix}$ with $N_{\rm pix} = 512$ and fixed pixel resolution $\epsilon = 2 h^{-1} \, {\rm Mpc}$ - the total volume $V = (1024 h^{-1} \, {\rm Mpc})^{3}$ is taken to be cubic but this is not necessary. We take a periodic domain so $\delta_{N_{\rm pix} + 1, j, k} = \delta_{1, j, k}$, $\delta_{i, N_{\rm pix} + 1, k} = \delta_{i, 1, k}$, $\delta_{i, j, N_{\rm pix} + 1} = \delta_{i,j, 1}$.  We normalise the field $\delta_{i,j,k} \to \delta_{i,j,k}/\sigma_{0}$, where $\sigma_{0}$ is the rms fluctuation of $\delta_{i,j,k}$ within the box. From this field we create triangulated surfaces of constant density $\delta = \nu$ by applying the method of marching tetrahedra \citep{tets91}. This algorithm runs through every grid point $(i,j,k)$ in the total volume, generating pixel cubes from the eight nearest neighbours $\delta_{i,j,k}, \delta_{i+1, j,k}$, $\delta_{i, j+1, k}$, $\delta_{i,j, k+1}$, $\delta_{i, j+1, k+1}$, $\delta_{i+1, j+1, k}$, $\delta_{i+1, j, k+1}$, $\delta_{i+1, j+1, k+1}$. This pixel cube is then decomposed into six non-overlapping, equal area tetrahedra. The density field is linearly interpolated along the edges of the tetrahedra to points at which $\delta = \nu$ and a triangulated surface is generated from these points. Details of the calculation can be found in appendix \ref{sec:3}. Once a surface of constant $\delta = \nu$ has been constructed, we extract the Minkowski functionals and tensors using their discretized approximations given in equations ($\ref{eq:w0}-\ref{eq:w3}$) and ($\ref{eq:mt021},\ref{eq:mt022}$). 

We first test the numerical algorithm by applying it to an isotropic Gaussian field. We generate discrete random fields $\delta_{i,j,k}$ in Fourier space drawn from a $\Lambda$CDM linear matter power spectrum, taking fiducial parameters shown in table \ref{tab:ii}. We smooth the field in Fourier space using Gaussian kernel $W(kR_{\rm G})=e^{-k^{2}R_{\rm G}^{2}/4}$ with $R_{\rm G} = 20 h^{-1} \, {\rm Mpc}$. We vary the threshold $-3 < \nu < 3$ using $N=40$ equi-spaced values. At each $\nu$ threshold we generate a triangulated surface of constant density that encloses all pixels satisfying $\delta_{i,j,k} > \nu$ -- this boundary defines the excursion set. We then use the properties of the surface to numerically reconstruct the scalar and tensor Minkowski functionals. In figure \ref{fig:8} we present the scalar Minkowski functionals $W_{0-3}$ as a function of threshold $\nu$, where blue points and error bars represent the mean and error on the mean of these statistics extracted numerically from $N_{\rm real} = 50$ realisations. The solid black line represents the theoretical expectation values given in equations ($\ref{eq:a5}-\ref{eq:a8}$). The algorithm accurately reproduces the expected scalar Minkowski functionals. This serves as a consistency check. 

We plot the Minkowski tensors $(W^{0,2}_{1})_{ij}$ (left panels) and $(W^{0,2}_{2})_{ij}$ (right panels) in figure \ref{fig:MT}. In the top panels we exhibit the diagonal components of these matrices $(ij)=(11,22,33)$ (red/yellow/blue points). We have shifted the points by $\delta \nu = \pm 0.03$ in the $x$-axis to distinguish them. The solid black lines are the theoretical expectation values for these fields, which exhibit sub-percent level agreement with the numerical output. We also plot the functions $\Delta W^{0,2}_{1}$ and $\Delta W^{0,2}_{2}$ defined as 

\begin{eqnarray}\label{eq:dw1} & & \Delta (W^{0,2}_{1})_{ij} =  { 3 (W^{0,2}_{1})_{ij} - \langle W_{1} \rangle \id_{ij} \over \langle W_{1} \rangle}  \\
\label{eq:dw2} & & \Delta (W^{0,2}_{2})_{ij} =  {3 (W^{0,2}_{2})_{ij} - \langle W_{2} \rangle \id_{ij} \over \langle W_{2} \rangle }     \end{eqnarray} 

\noindent These functions should be consistent with zero when $(W^{0,2}_{1}) \propto W_{1} \id$, $(W^{0,2}_{2}) \propto W_{2} \id$. 

In the middle panels we exhibit the off-diagonal components $(ij)=(12,13,23)$ (black, green, cyan points) divided by the mean of the diagonal components -- $\overline{W}^{0,2}_{1} \equiv \sum_{i=1}^{3} (W^{0,2}_{1})_{ii}/3$, $\overline{W}^{0,2}_{2} \equiv \sum_{i=1}^{3} (W^{0,2}_{2})_{ii}/3$. The off-diagonal elements should be consistent with zero for a Gaussian isotropic field. The points and error bars indicate the mean and error on the mean obtained from the $N_{\rm real} = 50$ realisations of the field. The off-diagonal components fluctuate around zero as expected. The error bars increase with increasing $|\nu|$ due to the smaller total surface area for large thresholds. This also occurs in the vicinity $\nu \sim 0$ for $W^{0,2}_{2}$. 

The top and middle panels of figure \ref{fig:MT} confirm the analytic results derived in section \ref{sec:2}. The ensemble averages of the matrices $W^{0,2}_{1}$, $W^{0,2}_{2}$ are proportional to products of the identity matrix and $W_{1}$, $W_{2}$ respectively for an isotropic, Gaussian random field. The statistics $\Delta W^{0,2}_{1}$ and $\Delta W^{0,2}_{2}$ provide a convenient measure of anisotropy -- these quantities should be identically zero for an isotropic field, regardless of cosmological parameters or power spectrum shape.

\section{Mean Shape of Excursion Regions}
\label{sec:5}

In the previous section we calculated the Minkowski functionals and generalised tensors for excursion sets defined over the entire volume of a three dimensional space. Further information can be extracted from the field by considering these statistics for every distinct sub-region contained within the excursion set. If we calculate the tensors $W^{0,2}_{1}$ and $W^{0,2}_{2}$ for every individual distinct surface that makes up the excursion set and also every individual hole, then we arrive at a distribution of tensors that provide information on the morphology of these constant density surfaces. As each sub-region will be oriented randomly, we diagonalise their $W^{0,2}_{1}$ and $W^{0,2}_{2}$ matrices and generate a set of eigenvalues $\lambda^{(1)}_{1,2,3}$ and $\lambda^{(2)}_{1,2,3}$. These parameters provide shape information for each structure, we define the mean ratio of these eigenvalues as 

\begin{eqnarray}\label{eq:bet1} & & \beta^{(1)}_{i} = \left\langle {\lambda^{(1)}_{i} \over \lambda^{(1)}_{1}} \right\rangle \\
\label{eq:bet2} & &\beta^{(2)}_{i} = \left\langle {\lambda^{(2)}_{i} \over \lambda^{(2)}_{1}} \right\rangle   \end{eqnarray}

\noindent where subscripts run over $i = 2,3$ and $\beta^{(1)}_{i} \le 1$, $\beta^{(2)}_{i} \le 1$ as we have ordered $\lambda^{(1)}_{1,2,3}$ and $\lambda^{(2)}_{1,2,3}$ in descending size. The sample mean, represented by angular brackets $\langle ... \rangle$, is for the combination of all connected regions and holes within the excursion set, for fixed threshold $\nu$. These quantities provide a measure of the average shape of iso-density structures as a function of threshold. $\beta^{(1,2)}_{i} = 1$ corresponds to a perfectly isotropic mean shape (sphere, cube) and $\beta^{(1,2)}_{i} < 1$ indicates either ellipticity or more general departure from isotropy.

To calculate $\lambda^{(1,2)}_{1,2,3}$ for every connected region and hole, we apply a type of friends-of-friends algorithm to the discretized density field $\delta_{i,j,k}$. We note that in each tetrahedron in our decomposition, all `in' states belong to the same excursion set sub-region (cf. figure \ref{fig:2}). Using this information, our algorithm passes through the volume systematically from one corner to the opposite, and each time it encounters a $(i,j,k)$ pixel that is `in' and not assigned to a particular excursion set sub-region, it is assigned to a new set. The algorithm then searches all $(i',j',k')$ pixels that this pixel is linked to via our tetrahedral decomposition; if these points are also `in' states then they are assigned to the same subset as the original point $(i,j,k)$. This procedure is repeated for all $(i', j', k')$ `in' states that have been found using the algorithm and then iteratively until no more linked `in' pixels are found. We then resume our original sweep through the box. The algorithm is repeated for `out' states of the field. Once we have categorized all `in' and `out' pixels according to the subset to which they belong, we can calculate the morphological and topological quantities of each sub-region by using the surface that encloses each of them.

In the bottom panels of figure \ref{fig:MT} we plot the mean eigenvalue ratios $\beta^{(1)}_{i}$, $\beta^{(2)}_{i}$ (left, right panels) for all connected regions and holes in the excursion set. In the large $|\nu|$ limit, these correspond to the shape of the field in the vicinity of peaks and troughs. One can predict the form of a Gaussian field in the vicinity of a peak \citep{Bardeen:1985tr}; the result is that the field will have the form of a triaxial ellipsoid. The calculation of the expectation value of the ellipticity $e_{\rm m}$ and prolateness $p_{\rm m}$ in the vicinity of a peak is presented in appendix \ref{app:2}. For an ellipse with parameters $e_{\rm m}$, $p_{\rm m}$ we can calculate the eigenvalues of the corresponding Minkowski tensors $W^{0,2}_{1}$ and $W^{0,2}_{2}$ using equations ($\ref{eq:intel1},\ref{eq:intel2}$); these analytic predictions are shown as solid black lines in the bottom panels of figure \ref{fig:MT}. We find good agreement between our numerical reconstruction of $\beta^{(1)}_{i}$, $\beta^{(2)}_{i}$ and the analytic prediction for large $|\nu|$ thresholds. However, the theoretical approximation does not accurately reproduce the details of the $\beta$ curves for $-3 < \nu < 3$ - this is not a surprising result as the theoretical prediction was constructed by expanding the density field as an ellipsoid in the vicinity of a peak. $\beta^{(1)}_{i}$ and $\beta^{(2)}_{i}$ are sensitive to the shape of the excursion set boundary, which will not be elliptical in general. In the large $|\nu|$ limit both $\beta^{(1)}_{i}$, $\beta^{(2)}_{i}$ are related to the mean axis lengths of the approximately elliptical bounding surface. The length of the axes vary with $\nu$ and show an overall tendency towards a spherical shape for large $|\nu|$.

The components of $W^{0,2}_{1}$ and $W^{0,2}_{2}$ have dimensions of area and length per unit volume respectively. Their eigenvalues, and hence $\beta^{(1)}_{i}$, $\beta^{(2)}_{i}$, are correlated and purely from the dimensionality of the matrices we expect $\beta^{(2)}_{\rm i} \sim \sqrt{\beta^{(1)}_{i}}$. This correspondence will not be exact, as the mean curvature $G_{2}$ acts as a weighting factor when comparing the two statistics.

When numerically reconstructing $\beta^{(1,2)}_{i}$ from a discrete density field, we apply a volume cut and only calculate the sample means ($\ref{eq:bet1}, \ref{eq:bet2}$) for excursion set regions and holes that have a volume greater than $V_{\rm cut} > 8 \epsilon^{3}$, where $\epsilon=2\, h^{-1} \, {\rm Mpc}$ is the resolution of the grid. Small scale peaks that are not resolved can generate artificial anisotropy and will spuriously decrease the $\beta^{(1,2)}_{i}$ functions. We discuss this issue further in appendix \ref{sec:dis_err}.


\section{Numerical Calculation of Minkowski Tensors for Anisotropic Gaussian Random Fields}
\label{sec:rsd}

In sections \ref{sec:2} and \ref{sec:4} we have predicted the ensemble averages of $W^{0,2}_{1}$ and $W^{0,2}_{2}$ for isotropic, Gaussian random fields and reconstructed them numerically from a discretely sampled density field. For these fields the Minkowski tensors are simple, being proportional to the identity matrix and scalar Minkowski functionals. The utility of these statistics is apparent when applying them to fields that are anisotropic or non-Gaussian. For an anisotropic yet Gaussian field, the Minkowski tensors will be characterized by matrices with unequal eigenvalues. Certain types of non-Gaussianity can potentially generate off-diagonal elements. In this work we focus on anisotropy and postpone an analysis of non-Gaussianity to the future.

We generate an anisotropic, Gaussian field by first creating an isotropic field $\delta_{i,j,k}$ in a $V = \left( 1024 h^{-1} \, {\rm Mpc}\right)^{3}$ box of resolution $\epsilon = 2 h^{-1} \, {\rm Mpc}$ drawn from a $\Lambda$CDM linear matter power spectrum with parameters defined in table \ref{tab:ii}, then applying a linear redshift space distortion operator, taking the line of sight arbitrarily as the $x_{3}$-axis. Specifically we generate $N_{\rm real} = 50$ realisations of a Gaussian random field drawn from the same $\Lambda$CDM power spectrum as in section \ref{sec:4}, and then apply the following transformation in Fourier space - 

\begin{equation} \delta^{\rm (rsd)}({\bf k}) = b\left( 1 + \Upsilon \mu_{\bf k}^{2}\right) \delta({\bf k}) \end{equation}

\noindent where $\mu_{\bf k}^{2} = k_{x_{3}}^{2}/k^{2}$ and $\Upsilon = f/b$ is the redshift space distortion parameter, where $b$ is the bias and $f \simeq \Omega_{\rm m}^{6/11}$ is the growth parameter\footnote{To avoid confusion with $\beta^{(1,2)}_{i}$, we have used $\Upsilon$ to denote the redshift space distortion parameter rather than the standard notation $\beta=f/b$.}. For simplicity we use the plane-parallel approximation, which reduces the effect of linear redshift space distortion to the Kaiser effect corresponding to an amplitude shift $\left( 1 + \Upsilon \mu_{\bf k}^{2}\right)$ in Fourier space. This field is then smoothed in Fourier space using a Gaussian kernel of width $R_{\rm G} = 20 h^{-1} \, {\rm Mpc}$. We calculate the Minkowski tensors for both the isotropic case $\Upsilon=0$, and for a redshift space distorted field $\Upsilon \neq 0$. Since we are using the matter density field directly, as opposed to a biased tracer such as galaxies or halos, we fix $b=1$ and $\Upsilon = \Omega_{\rm mat}^{6/11}$. The redshift-dependence of $\Upsilon$ will result in a redshift-dependent signal in $W^{0,2}_{1}$, $W^{0,2}_{2}$. In this section we generate fields at $z=0$.

\begin{figure*}[!b]
  \includegraphics[width=1.0\textwidth]{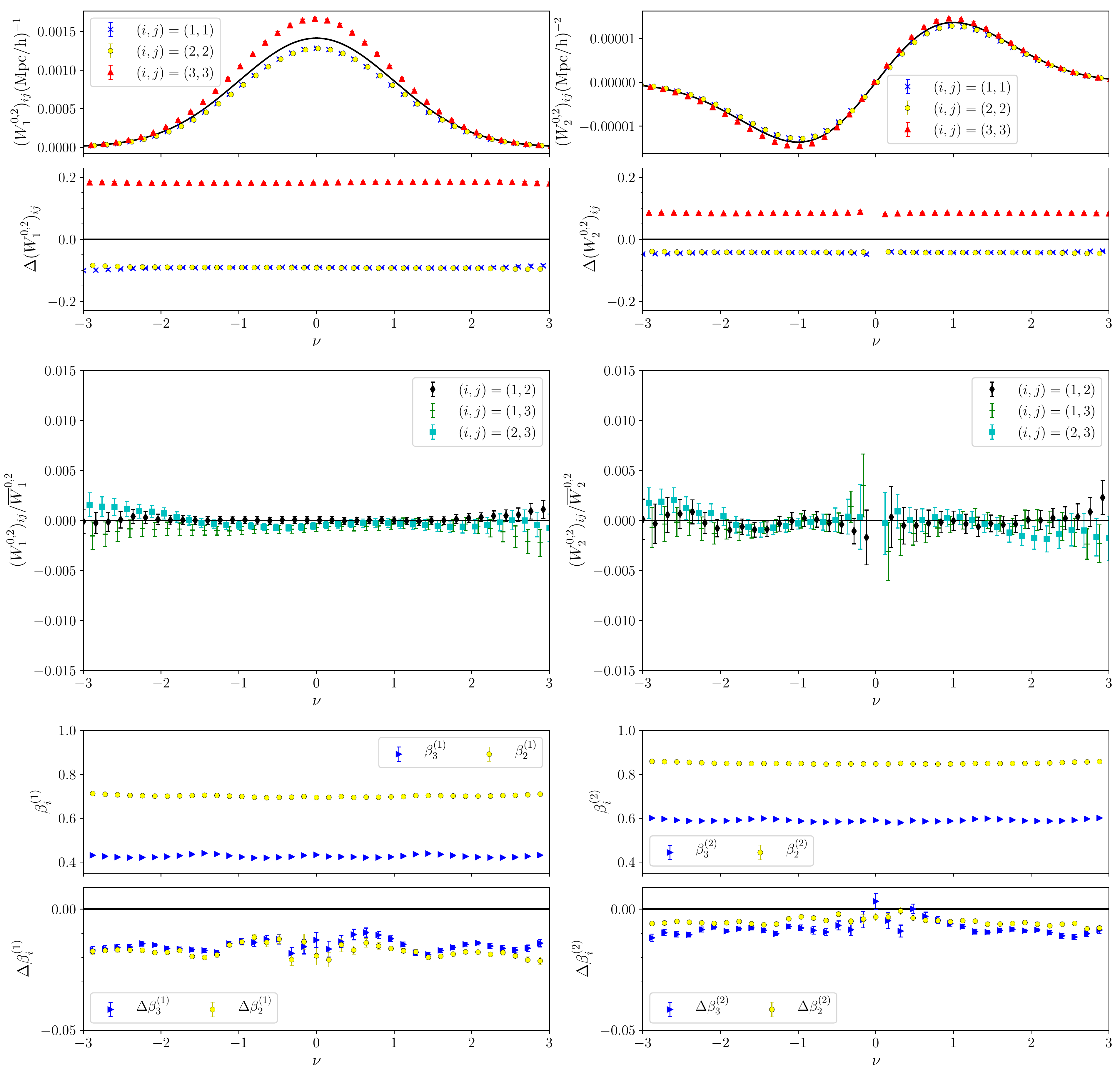} 
  \caption{The Minkowski tensors $W^{0,2}_{1}$ (left panels), $W^{0,2}_{2}$ (right panels) for the anisotropic Gaussian field described in section \ref{sec:rsd}. [Top panels] diagonal components of $(W^{0,2}_{1})_{ij}$, $(W^{0,2}_{2})_{ij}$, $(i,j)=(1,1),(2,2),(3,3)$ (blue,yellow,red points) and the theoretical expectation value for the corresponding isotropic field (black solid line). The components of $\Delta (W^{0,2}_{1})_{ij}$ and $\Delta (W^{0,2}_{2})_{ij}$ are no longer consistent with zero, showing a constant amplitude shift due to the anisotropic signal. [Middle panels] off-diagonal components of $(W^{0,2}_{1})_{ij}$, $(W^{0,2}_{2})_{ij}$ divided by the average of the diagonal components $\overline{W}^{0,2}_{1}$, $\overline{W}^{0,2}_{2}$. These quantities are consistent with zero as expected for a Gaussian field. [Bottom panels] $\beta^{(1)}_{i}$ (left), $\beta^{(2)}_{i}$ (right) for the anisotropic field. $\Delta \beta^{(1,2)}_{i}$ is the difference between $\beta^{(1,2)}_{i}$ measured for the anisotropic $\Upsilon = \Omega_{\rm mat}^{6/11}$ and isotropic $\Upsilon=0$ fields. Excursion sets in the isotropic field are more spherical, characterised by higher $\beta^{(1,2)}_{i}$ values. }
  \label{fig:MT_RSD}
\end{figure*}

In figure \ref{fig:MT_RSD} we exhibit the diagonal components of the matrices $W^{0,2}_{1}$ (top left panel) and $W^{0,2}_{2}$ (top right panel). The solid black line is the theoretical prediction for an isotropic field, that is for $\Upsilon = 0$. The yellow/blue/red points represent the $(i,j)=(1,1), (2,2) (3,3)$ directions. The effect of linear redshift space distortion is to modify the diagonal elements, with the line of sight component $x_{3}$ enhanced relative to the perpendicular $x_{1,2}$ components. $\Delta (W^{0,2}_{1})_{ij}$ and $\Delta (W^{0,2}_{2})_{ij}$ are plotted in the second set of panels. These statistics, which are consistent with zero for an isotropic field, present a constant shift for all $-3 < \nu < 3$ thresholds. Hence the effect of adding linear redshift space distortion is a constant amplitude shift between the diagonal elements. The matrices are no longer proportional to the identity matrix, and the signal is more pronounced in the statistic $W^{0,2}_{1}$. 

In the third set of panels we exhibit the off-diagonal components of the matrices, divided by the average value of the diagonal elements. Black, green, cyan points correspond to $(i,j)=(1,2),(1,3),(2,3)$ respectively. The off-diagonal elements are all consistent with zero, as we expect. The linear redshift space distortion operator in the plane-parallel approximation corresponds to a directional dependent cofactor in Fourier space -- hence if we take a Gaussian field and apply this operator we expect the resulting field to remain Gaussian. This is reflected in the off-diagonal components of the Minkowski Tensors, which are consistent with zero. Note that we are aligning the coordinate system with the anisotropic signal in the $x_{3}$ axis. For an arbitrary coordinate system one should diagonalise the matrix, and any anisotropic signal would manifest as inequality between eigenvalues.

In the lower panels we plot the mean shape of connected regions and holes within the excursion set for the redshift space distorted field, characterised by $\beta^{(1)}_{i}$ (left panel) and $\beta^{(2)}_{i}$ (right panel). We also plot $\Delta \beta^{(1)}_{i}$ and $\Delta \beta^{(2)}_{i}$, which are defined as the difference between the functions $\beta^{(1)}_{i}$, $\beta^{(2)}_{i}$ as measured in redshift and real space

\begin{equation} \Delta \beta^{(1,2)}_{i} = \beta^{(1,2)}_{i, {\rm rsd}}  - \beta^{(1,2)}_{i, {\rm real}} .   \end{equation} 

\noindent One can observe a small $\sim 1\%$ decrease in $\beta^{(1,2)}_{i}$ in redshift space relative to real space; this is due to the elongation of structures along the line of sight as a result of coherent in-fall/outfall of matter into over/under-densities. The effect is roughly constant as a function of $\nu$, and is more pronounced for the Minkowski tensor $W^{0,2}_{1}$. Although the effect of redshift space distortion on each individual excursion set region is very small, the signal is cumulative in the sense that each will be distorted in the same direction. This cumulative signal generates the large effect on $\Delta W^{0,2}_{1}$, $\Delta W^{0,2}_{2}$ observed in the second set of panels.

Figure \ref{fig:MT_RSD} indicates that the redshift space parameter $\Upsilon=f/b$ can be constrained by measuring the diagonal elements of $W^{0,2}_{1}$, $W^{0,2}_{2}$. We attempt to quantify the sensitivity of these statistics to $\Upsilon$ by varying $\Omega_{\rm mat}$ and hence $\Upsilon \simeq \Omega_{\rm mat}^{6/11}$. We generate Gaussian, anisotropic fields for three different values $\Omega_{\rm mat} = (0.21, 0.26, 0.31)$ and reconstruct the Minkowski tensors in each case. A larger $\Omega_{\rm mat}$ will produce a more pronounced anisotropic signal, and we study the sensitivity of the amplitude shift in $\Delta (W^{0,2}_{1})_{ij}$, $\Delta (W^{0,2}_{1})_{ij}$ to this parameter. 

\begin{figure*}[!b]
  \includegraphics[width=1.0\textwidth]{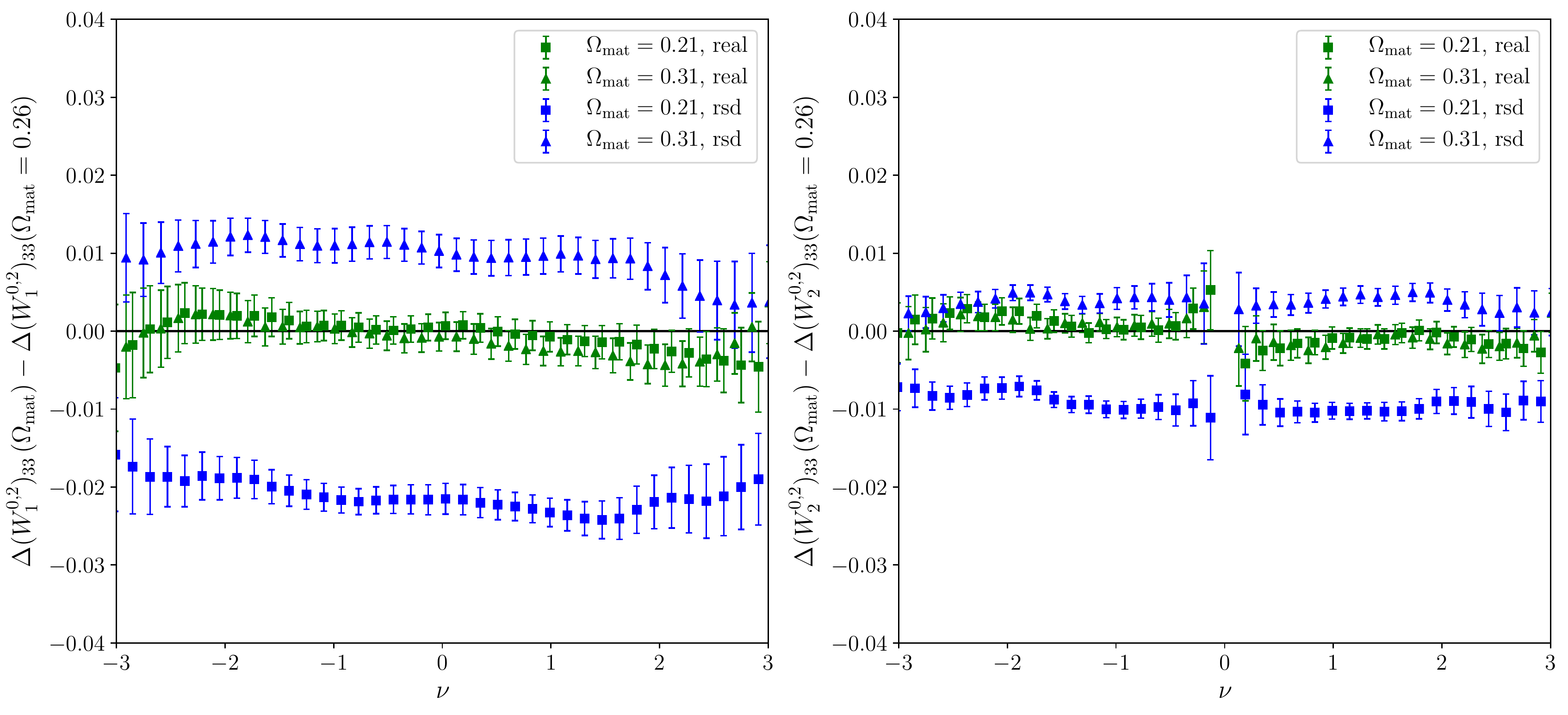} 
  \caption{[Left panel] The difference between $\Delta (W^{0,2}_{1})_{33}$ measured for fields generated with $\Omega_{\rm mat}=0.21$ (squares) and $\Omega_{\rm mat} = 0.31$ (triangles) and the `fiducial' cosmology $\Omega_{\rm mat} = 0.26$. The green data points are for an isotropic field and are consistent with zero. For an anisotropic field (blue points) $\Delta (W^{0,2}_{1})_{33}$ is non-zero and sensitive to $\Omega_{\rm mat}$. [Right panel] As in the left panel but using the statistic $\Delta (W^{0,2}_{2})_{33}$.  }
  \label{fig:alpha_comp_real}
\end{figure*}

In figure \ref{fig:alpha_comp_real} we plot the differences $\Delta (W^{0,2}_{1})_{33}(\Omega_{\rm mat}) - \Delta (W^{0,2}_{1})_{33}(\Omega_{\rm mat}=0.26)$ (left panel), $\Delta (W^{0,2}_{2})_{33}(\Omega_{\rm mat}) = \Delta (W^{0,2}_{2})_{33}(\Omega_{\rm mat}=0.26)$ (right panel) extracted from $N_{\rm real}=50$ realisations of Gaussian fields. We only plot the $(ij)=(33)$ components for clarity. The $(ij)=(11),(22)$ components exhibit similar behaviour. The green points correspond to the isotropic field $\Upsilon=0$ and blue points the linear redshift space distorted field $\Upsilon = \Omega_{\rm mat}^{6/11}$. The square/triangle data points correspond to $\Omega_{\rm mat} = 0.21$, $\Omega_{\rm mat} = 0.31$ respectively. 

For an isotropic field, $\Delta (W^{0,2}_{1})_{ij}(\Omega_{\rm mat})$ and $\Delta (W^{0,2}_{2})_{ij}(\Omega_{\rm mat})$ are consistent with zero for all $\Omega_{\rm mat}$, and hence linear combinations should also be zero (cf green points). For a redshift space distorted field, the matrices $\Delta (W^{0,2}_{1})_{ij}(\Omega_{\rm mat})$, $\Delta (W^{0,2}_{2})_{ij}(\Omega_{\rm mat})$ are no longer consistent with zero -- the amplitude shift observed in figure \ref{fig:MT_RSD} will depend on $\Omega_{\rm mat}$. In figure \ref{fig:alpha_comp_real} we find that an order $\sim {\cal O}(10\%)$ variation of $\Upsilon$ will generate an order $\sim {\cal O}(2\%)$ variation in $\Delta (W^{0,2}_{1})_{ij}(\Omega_{\rm mat})$ and a smaller variation in $\Delta (W^{0,2}_{2})_{ij}(\Omega_{\rm mat})$. Although the signal is small, we measure the statistics over multiple thresholds $\nu$ which will increase the statistical significance of the amplitude shift.

For isotropic fields, the diagonal components of $W^{0,2}_{1}$ and $W^{0,2}_{2}$ are sensitive to the value of $\Omega_{\rm mat}$, however the matrix is always proportional to the identity matrix. It follows that $\Delta W^{0,2}_{1}$, $\Delta W^{0,2}_{2}$ should be insensitive to $\Omega_{\rm mat}$. For redshift space distorted fields, the diagonal elements of $W^{0,2}_{1}$ or $W^{0,2}_{2}$ are no longer equal, and the magnitude of the difference between them will depend on $\Upsilon$. Schematically we can write 

\begin{eqnarray} & & W^{0,2}_{1} \propto W_{1} \times {\rm diag} \{1,1,c(\Upsilon) \} \\
 & & W^{0,2}_{2} \propto W_{2} \times {\rm diag}\{ 1,1,c'(\Upsilon)\} \end{eqnarray}

\noindent where $c(\Upsilon), c'(\Upsilon) > 1$ are functions of $\Upsilon$ and we have chosen a coordinate system in which the anisotropic signal is aligned with the $x_{3}$ axis. The statistics $\Delta W^{0,2}_{1}$, $\Delta W^{0,2}_{2}$ will be sensitive to $c(\Upsilon), c'(\Upsilon)$ and hence $\Omega_{\rm mat}$.

Recall that we are using the matter field directly rather than a biased tracer, so are taking $b=1$. When applying the statistics to data, measuring $\Delta W^{0,2}_{1}$, $\Delta W^{0,2}_{2}$ will allow us to constrain $\Upsilon = f/b$. By measuring the genus $W_{3}$ of the three dimensional density field one can obtain a constraint on $\Omega_{\rm mat}$ that is relatively insensitive to galaxy bias and redshift space distortion. It follows that combinations of measurements of $W_{3}$ and $\Delta W^{0,2}_{1}$, $\Delta W^{0,2}_{2}$ can potentially provide simultaneous constraints on $\Omega_{\rm mat}$ and galaxy bias $b$.

\section{Summary} 
\label{sec:sum}

The purpose of this work has been to introduce the Minkowski tensors for three dimensional fields, as a generalisation of the standard scalar Minkowski functionals. Initially focusing on isotropic Gaussian random fields, we calculated the ensemble expectation value of the translation invariant statistics $W^{0,2}_{1}$ and $W^{0,2}_{2}$. In the appendix we provide an algorithm that generates closed triangulated iso-field surfaces, from which one can extract these statistics numerically from a discrete density field. We found close agreement between the theoretical expectation value and numerical reconstruction of the Minkowski functionals and tensors. 
 
For an isotropic, Gaussian random field the Minkowski tensors are proportional to the product of the scalar Minkowski functionals and the identity matrix. As such, $W^{0,2}_{1}$ and $W^{0,2}_{2}$ do not contain additional information relative to $W_{1}$ and $W_{2}$ respectively. However, departures from either Gaussianity or isotropy can potentially modify $W^{0,2}_{1}$ and $W^{0,2}_{2}$, either by generating off-diagonal terms or introducing inequality between diagonal elements. 

We applied our numerical algorithm to a Gaussian but anisotropic field, simulating a plane-parallel, linearly redshift space distorted density field. The effect of redshift space distortion will be to distort the shape of structures along the line of sight, generating a preferred direction in the normals of the excursion sets. As this effect is cumulative, we find a large $10-20\%$ amplitude shift in the $\Delta W^{0,2}_{1}$, $\Delta W^{0,2}_{2}$ statistics relative to an isotropic density field. This signal can be used to constrain the redshift space distortion parameter $\Upsilon = f/b$. 

In this work we have focused on linear density fields and the Kaiser approximation, which represents coherent in-fall of mass into overdensities. We expect $W^{0,2}_{1}$, $W^{0,2}_{2}$ will also be sensitive to non-linear anisotropies arising from the Finger of God effect. Conversely, we do not expect the statistics to be sensitive to non-linear gravitational collapse, as this process remains statistically isotropic on all scales. Hence these statistics will provide a measure of both the linear and non-linear velocity perturbation that is independent of non-linear behaviour of the density field $\delta$. A detailed analysis of non-Gaussianity will be pursued elsewhere, with an application of the Minkowski tensors to the late time gravitationally evolved matter density field. The eigenvalues of $W^{0,2}_{1}$, $W^{0,2}_{2}$ are independent measures of non-Gaussianity of a field and contain complementary information compared to the scalar Minkowski functionals.

\acknowledgements{\noindent Acknowledgements - The authors thank the Korea Institute for Advanced Study for providing computing resources (KIAS Center for Advanced Computation Linux Cluster System) for this work.}

\newpage

\appendix

\section{A. Generating the bounding surface of an excursion set} 
\label{sec:3}

\begin{figure}
 \centering
  \includegraphics[width=0.45\textwidth]{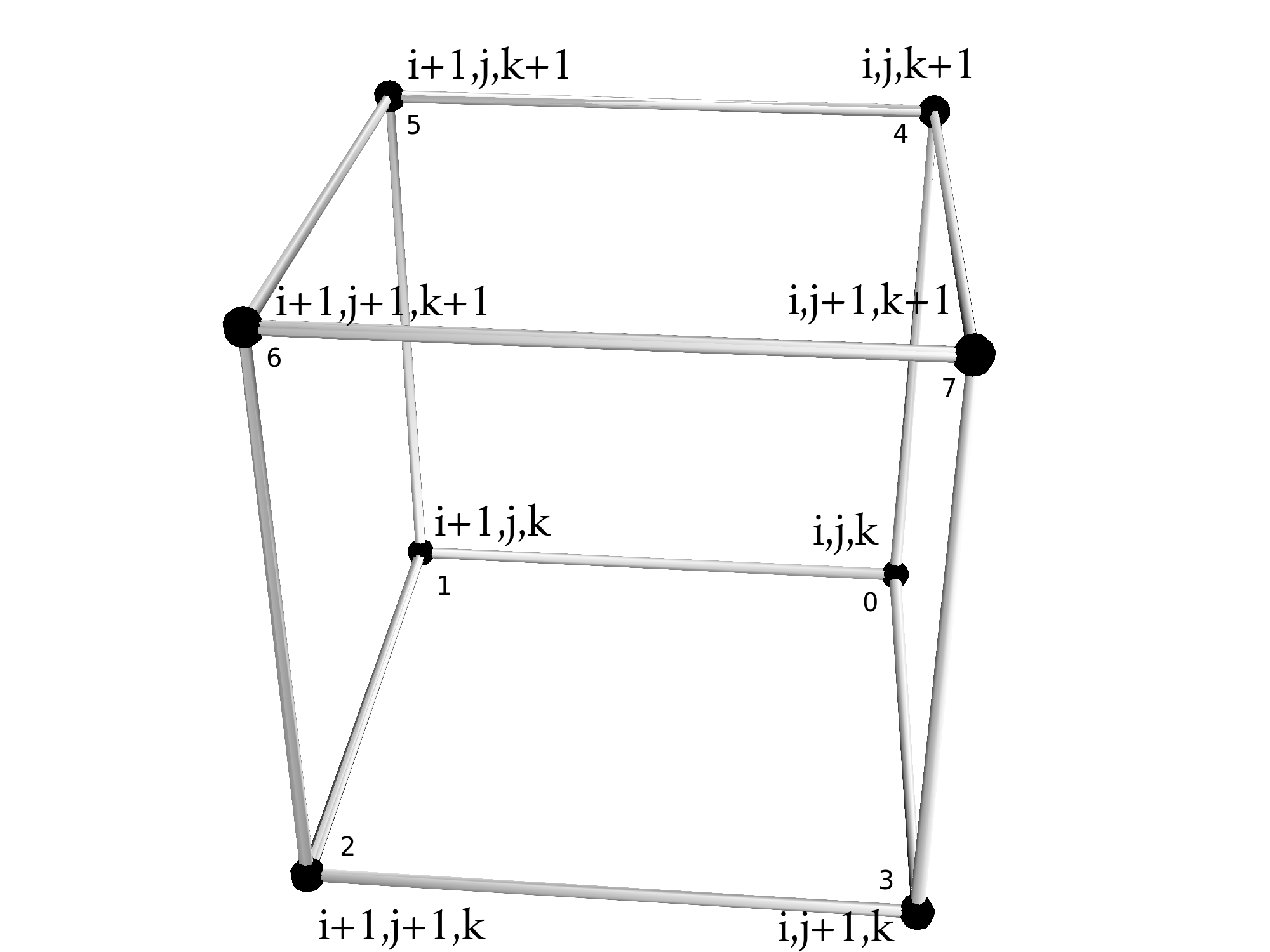}
  \caption{A pixel cube constructed from the eight nearest neighbours of the discretized density field $\delta_{i,j,k}$. We label the cube vertices $0-7$ as shown.}
  \label{fig:1}
\end{figure}

In the appendix we elucidate the numerical algorithm that we use to generate iso-density surfaces. Our starting point is a discretized three dimensional field $\delta_{i,j,k}$ on a regular lattice. We take $1 \le i,j,k \le N_{\rm pix}$ and constant pixel resolution $\epsilon$. We will study periodic density fields and so we take $\delta_{N_{\rm pix} + 1, j, k} = \delta_{1, j, k}$, $\delta_{i, N_{\rm pix} + 1, k} = \delta_{i, 1, k}$, $\delta_{i, j, N_{\rm pix} + 1} = \delta_{i,j, 1}$. We apply a density threshold $\nu$ and define pixels as `in' the excursion set if they satisfy $\delta_{i,j,k} > \nu$ and `out' if $\delta_{i,j,k} < \nu$. Our goal is to generate a surface of constant density $\delta = \nu$ that encloses the `in' pixels. We use the method of marching tetrahedra to construct the bounding surface.

\begin{figure}
  \centering
  \includegraphics[width=0.24\textwidth]{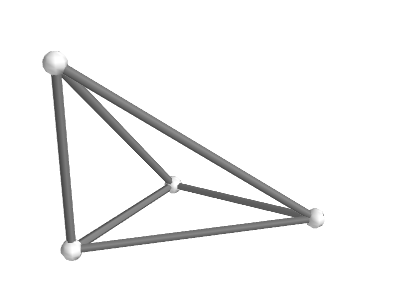}
  \includegraphics[width=0.24\textwidth]{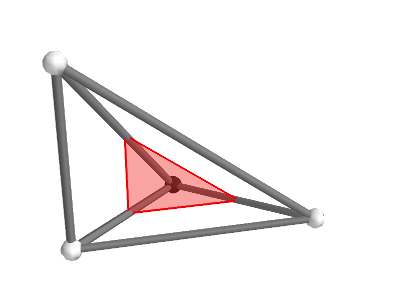}
  \includegraphics[width=0.24\textwidth]{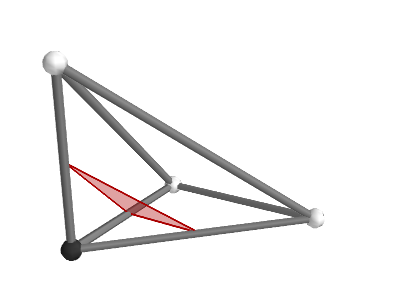}
  \includegraphics[width=0.24\textwidth]{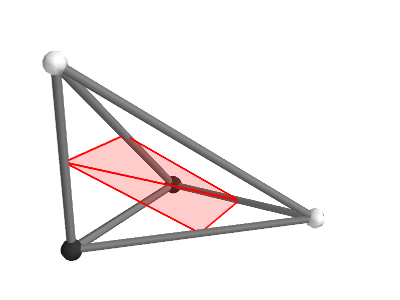}\\
  \includegraphics[width=0.24\textwidth]{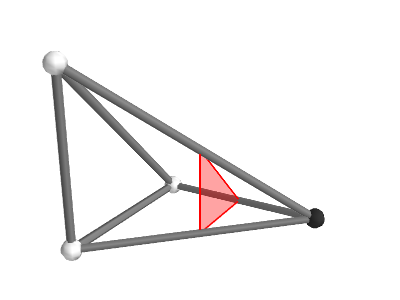}
  \includegraphics[width=0.24\textwidth]{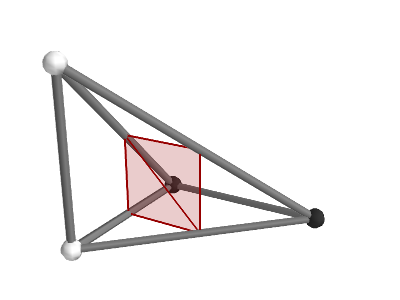}
  \includegraphics[width=0.24\textwidth]{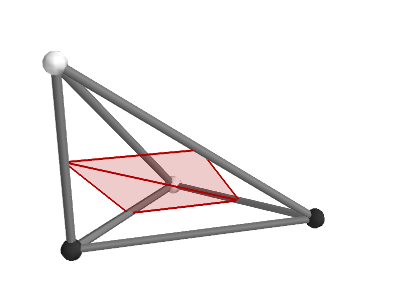}
  \includegraphics[width=0.24\textwidth]{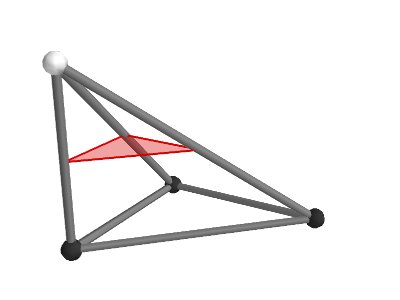}\\
  \includegraphics[width=0.24\textwidth]{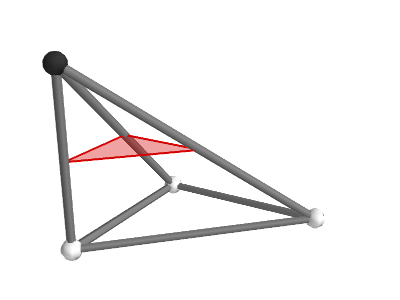}
  \includegraphics[width=0.24\textwidth]{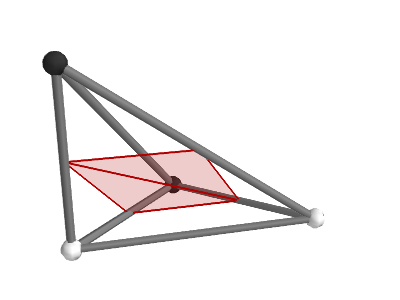}
  \includegraphics[width=0.24\textwidth]{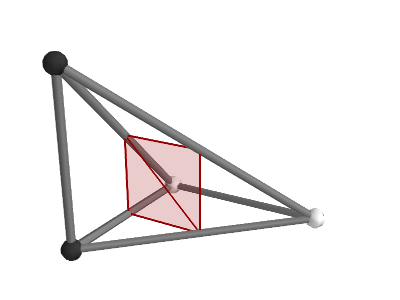}
  \includegraphics[width=0.24\textwidth]{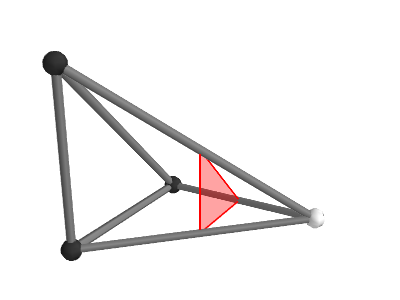}\\
  \includegraphics[width=0.24\textwidth]{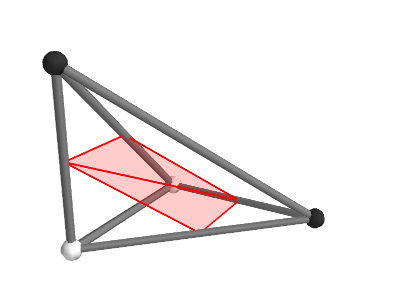}
  \includegraphics[width=0.24\textwidth]{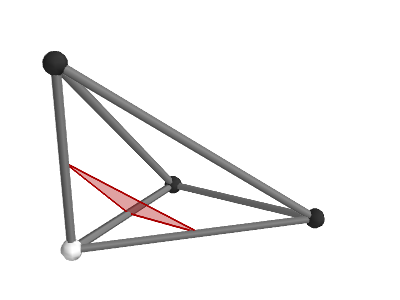}
  \includegraphics[width=0.24\textwidth]{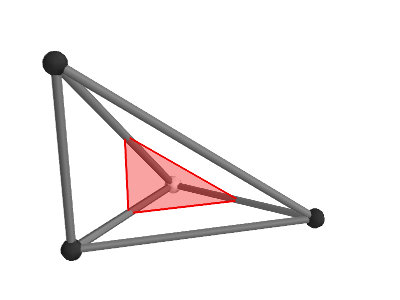}
  \includegraphics[width=0.24\textwidth]{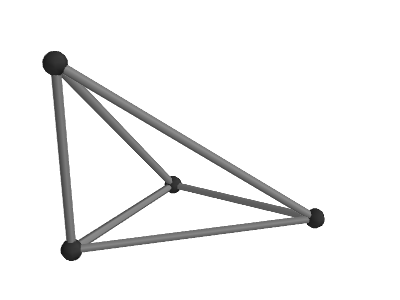}\\
  \caption{After decomposing a pixel cube into six tetrahedra (as described in the text), each tetrahedron can occupy one of sixteen possible states depending on whether its vertices are `in' ($\delta_{i,j,k} > \nu$, black points) or `out' ($\delta_{i,j,k} < \nu$, white points) of the excursion set. A bounding surface separating the in and out states is constructed by linearly interpolating along the edges of the tetrahedron to the point $\delta = \nu$, where $\nu$ is the density threshold. This procedure generates the vertices of a triangulation, displayed in red in the figure.   }
  \label{fig:2}
\end{figure}

The method of marching tetrahedra is similar in spirit to marching squares in two dimensions. We sweep through the total volume systematically over $i ,j, k$ dimensions, forming cubes of adjacent vertices $\delta_{i,j,k}, \delta_{i+1, j,k}$, $\delta_{i, j+1, k}$, $\delta_{i,j, k+1}$, $\delta_{i, j+1, k+1}$, $\delta_{i+1, j+1, k}$, $\delta_{i+1, j, k+1}$, $\delta_{i+1, j+1, k+1}$ - we exhibit one such cube in figure \ref{fig:1}. We then further subdivide each cube into six non-overlapping, equal area tetrahedra. All tetrahedra share a single edge, which is a major interior diagonal of the cube. We note that there is no single unique decomposition of a cube into tetrahedra, as there are four major diagonals. Using the vertex labels in figure  \ref{fig:1}, the tetrahedral decomposition of the cube can be defined in four ways $(a-d)$ as shown in table \ref{tab:0}. In this work we present results using decomposition (a), however when calculating the Minkowski functionals and their generalisations we have checked that all four yield consistent results.  


\begin{table}[!b]
\begin{center}
\caption{\label{tab:0} }
 \begin{tabular}{c | c | c | c}
 (a) & (b) & (c) & (d)  \\ [0.5ex] 
 \hline
 (0, 6, 2, 1) & (4, 2, 6, 5) &  (3, 5, 6, 7) & (7, 1, 2, 3)  \\ 
 (0, 6, 2, 3) & (4, 2, 6, 7) &  (3, 5, 6, 2) & (7, 1, 2, 6) \\ 
 (0, 6, 5, 4) & (4, 2, 1, 0) &  (3, 5, 1, 2) & (7, 1, 5, 6) \\ 
 (0, 6, 7, 4) & (4, 2, 3, 0) &  (3, 5, 1, 0) & (7, 1, 5, 4) \\
 (0, 6, 7, 3) & (4, 2, 3, 7) &  (3, 5, 4, 7) & (7, 1, 0, 3) \\
 (0, 6, 5, 1) & (4, 2, 1, 5) &  (3, 5, 4, 0) & (7, 1, 0, 4) \\
\end{tabular}
\end{center} 
Four possible tetrahedral decompositions of the pixel cubes (a-d), using index labels shown in figure \ref{fig:1}. 
\end{table}

Once an individual cube has been decomposed, we consider each of its six tetrahedra in turn. They possess four vertices, and each can be in/out of the excursion region. Therefore there are $2^{4} = 16$ possible distinct states of the tetrahedron. We exhibit the sixteen states in figure \ref{fig:2}. The solid points are vertices of the cube, with white representing `out' states and black `in'. A triangulation is constructed by linearly interpolating the density along any edge of the tetrahedron that joins an `in' $\delta > \nu$ (black points) and `out' $\delta < \nu$ (white points) state, to the point at which $\delta = \nu$. The triangulations are exhibited in red in figure \ref{fig:2}. Once we have generated the triangle vertices, we can calculate the total area of the triangles, the volume they enclose and the normals to the triangulated bounding surface, which always point externally to the enclosed `in' states.

\begin{figure}
  \centering
  \includegraphics[width=0.45\textwidth]{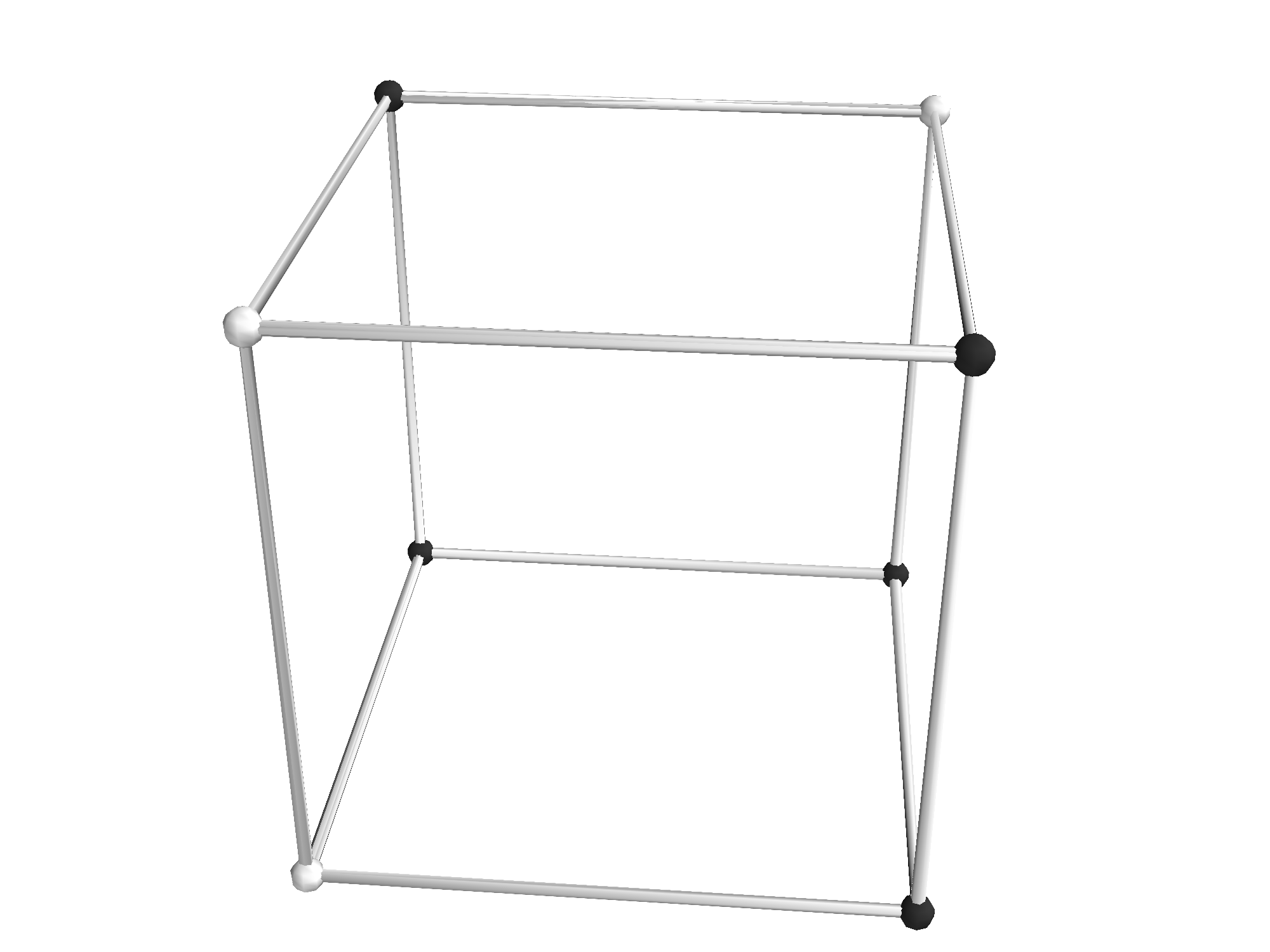}
  \includegraphics[width=0.45\textwidth]{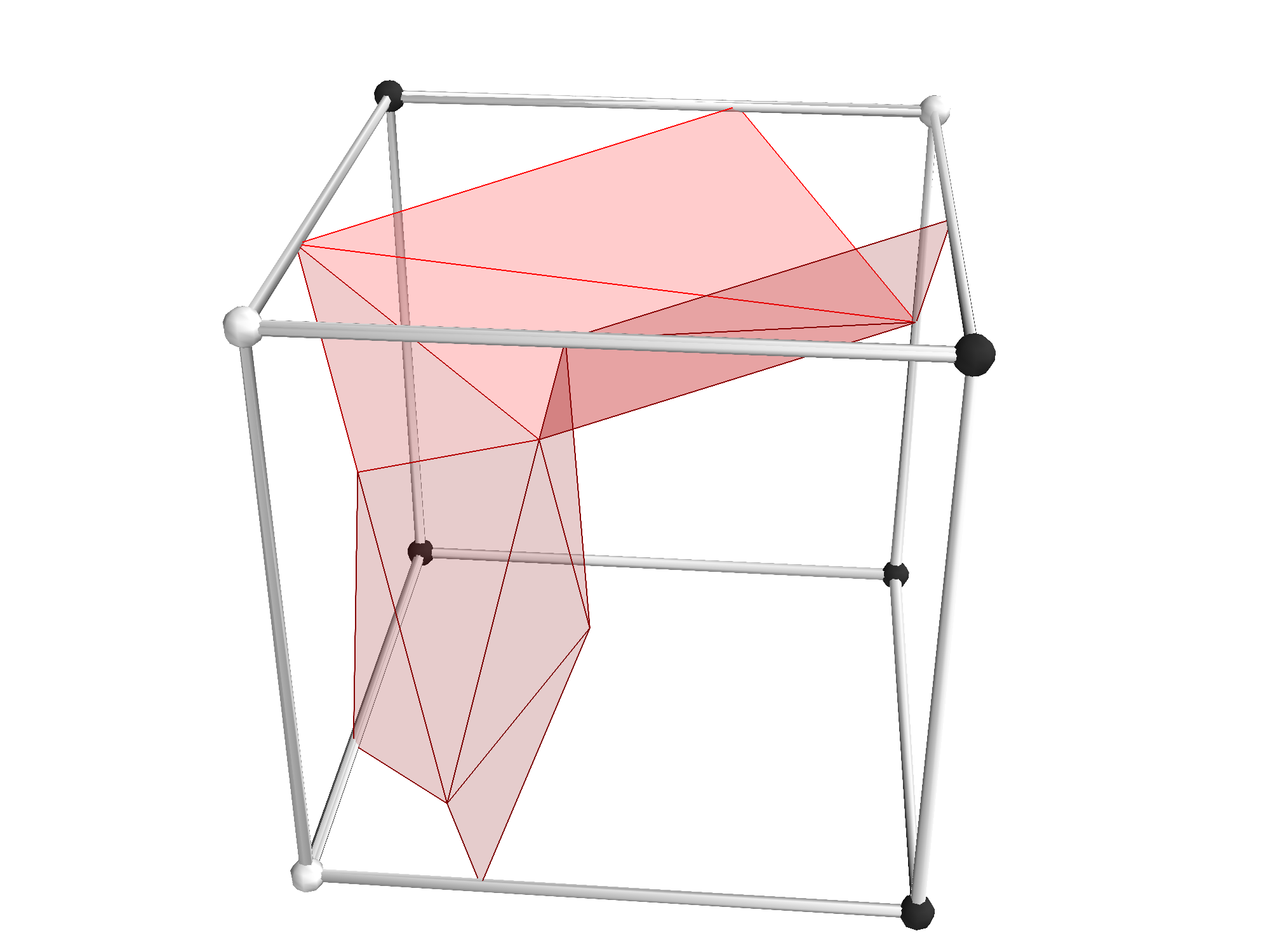}\\
  \includegraphics[width=0.32\textwidth]{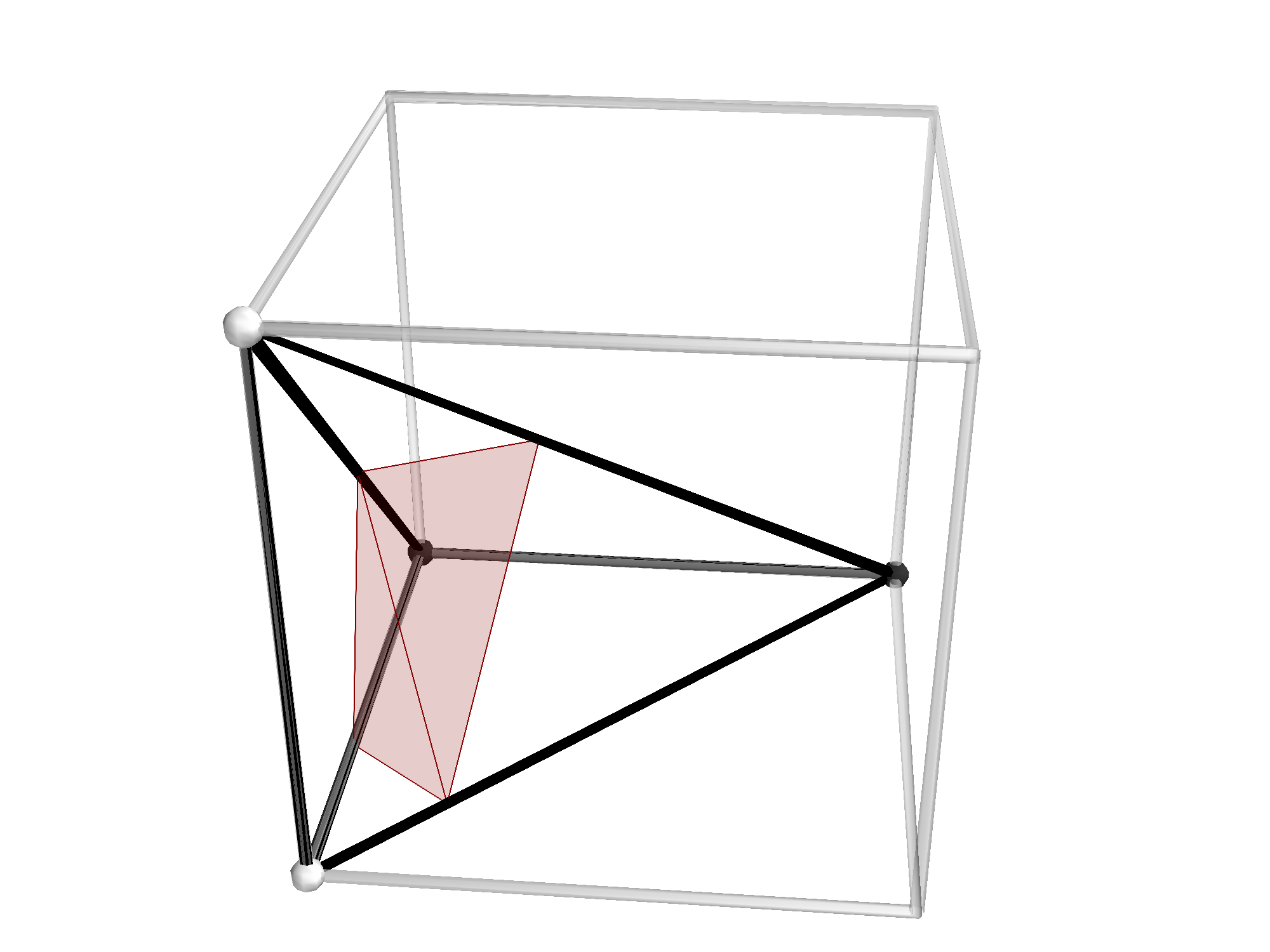}
  \includegraphics[width=0.32\textwidth]{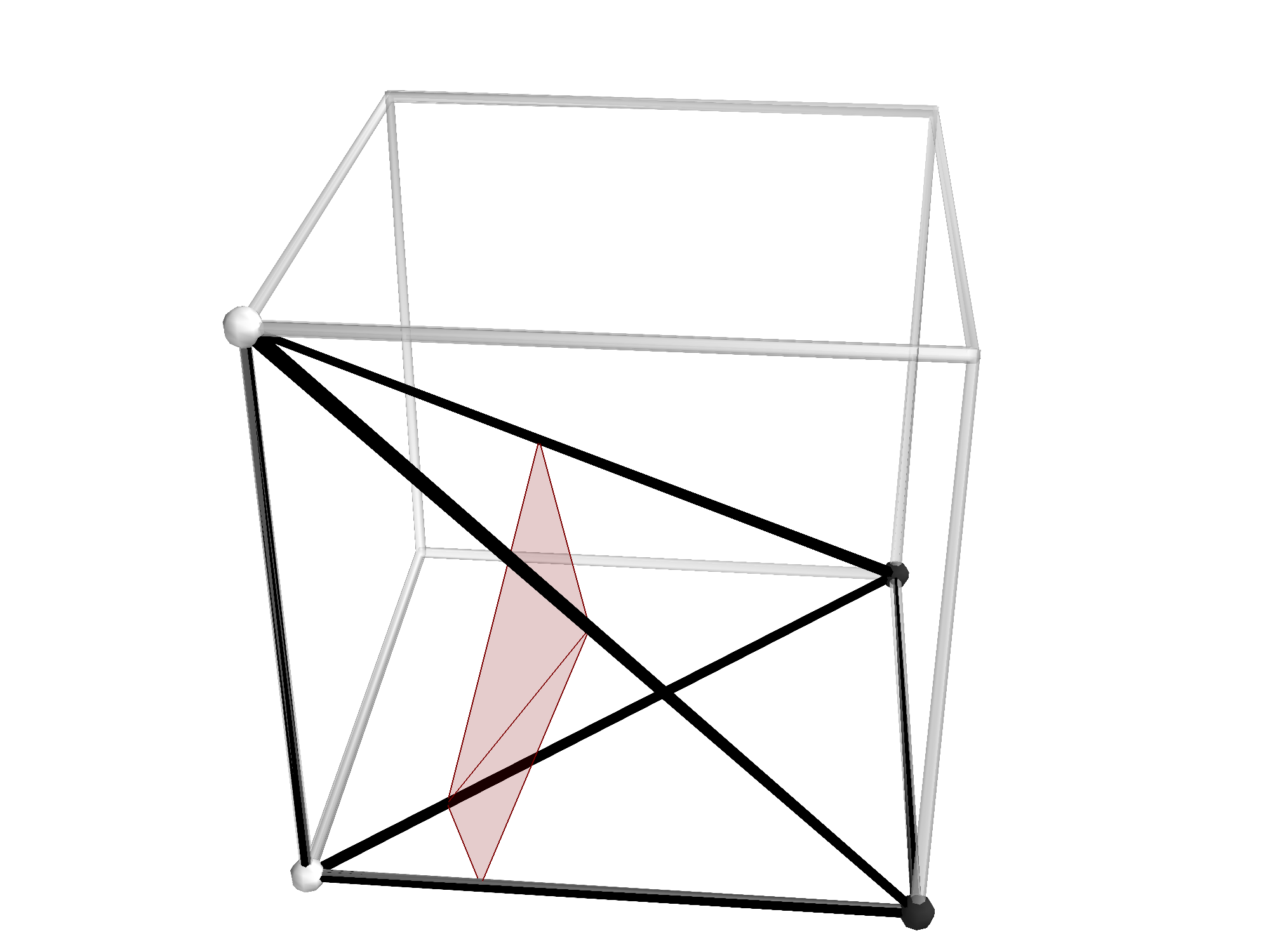}
  \includegraphics[width=0.32\textwidth]{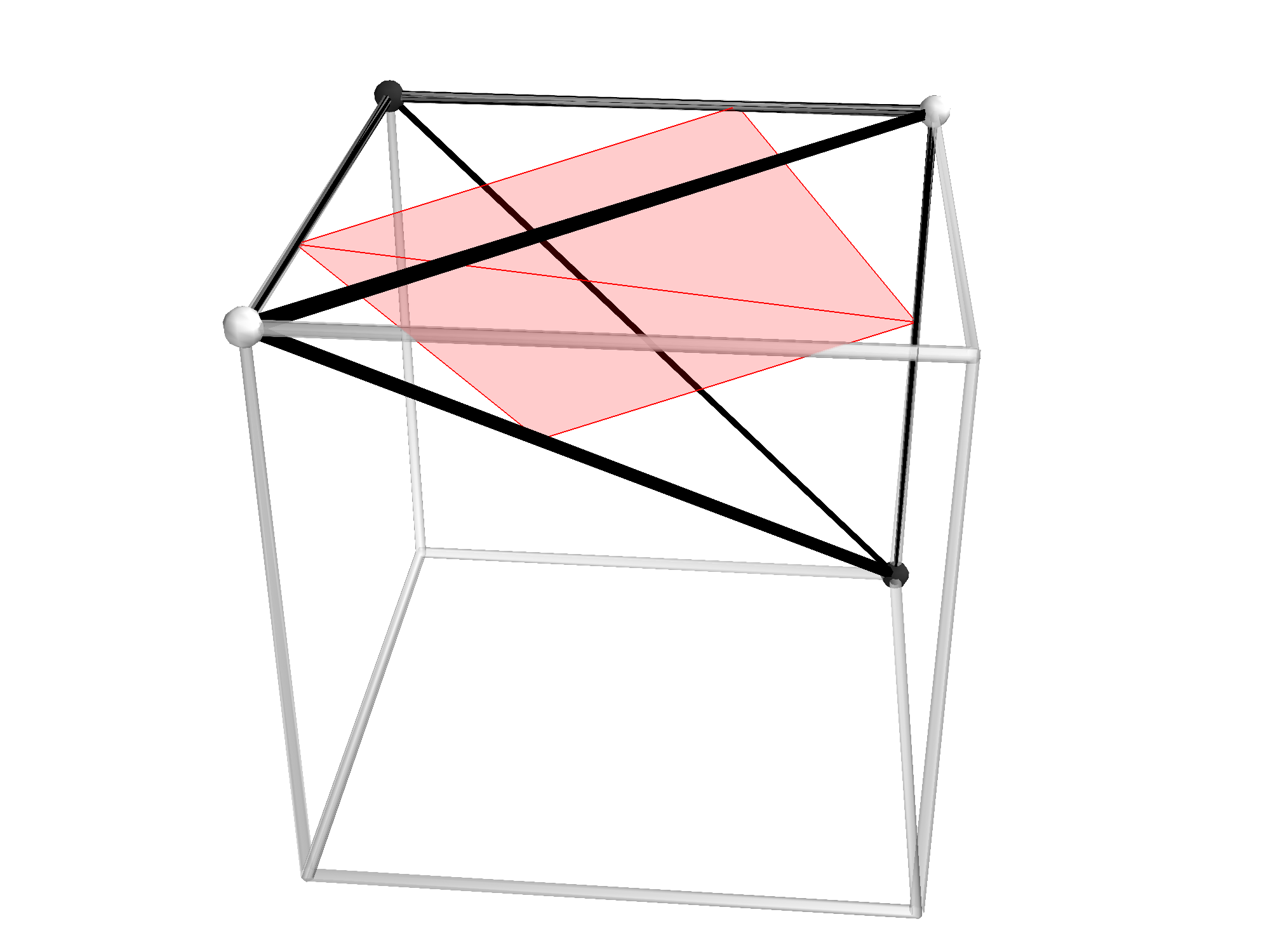}\\
  \includegraphics[width=0.32\textwidth]{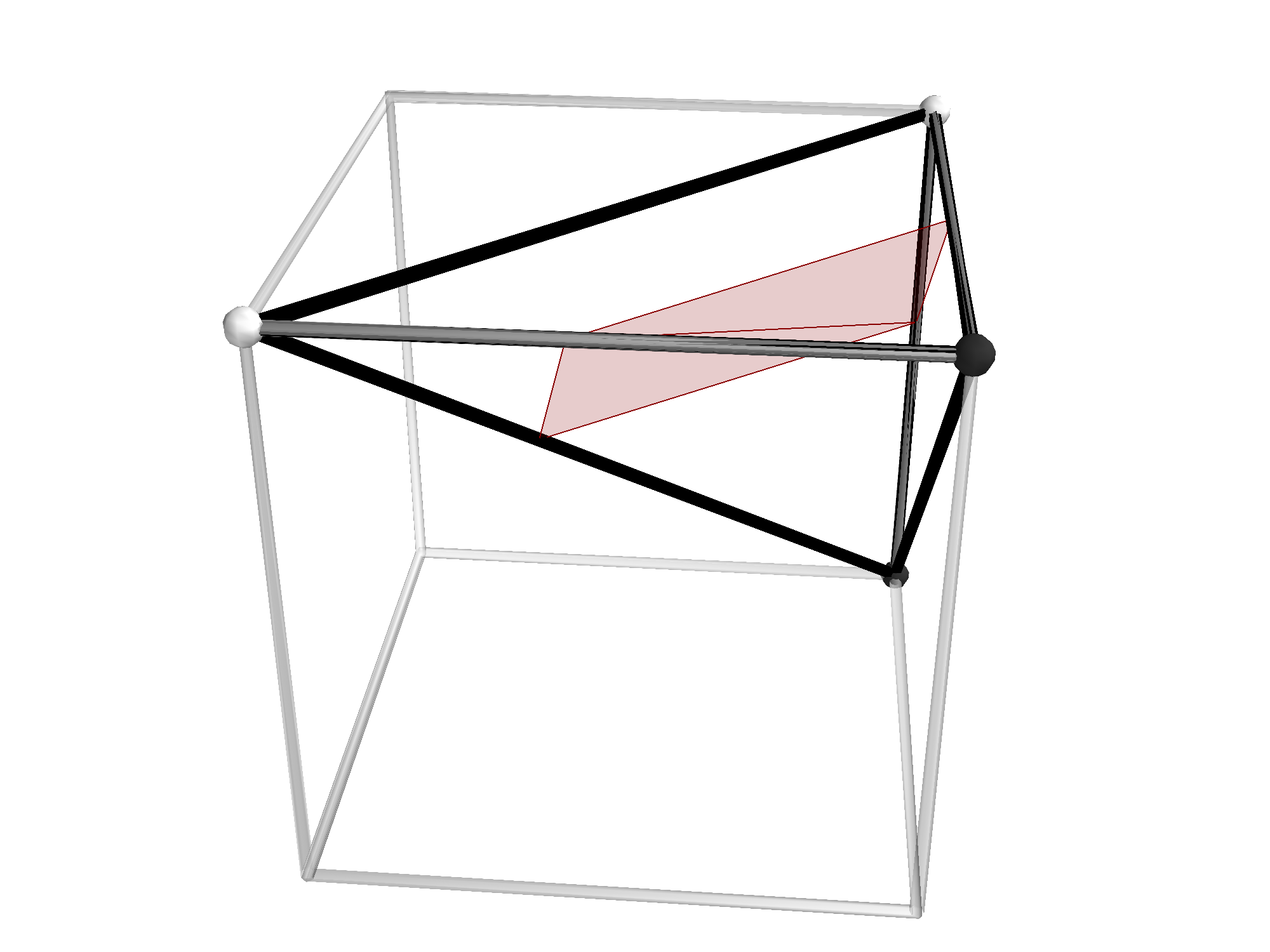}
  \includegraphics[width=0.32\textwidth]{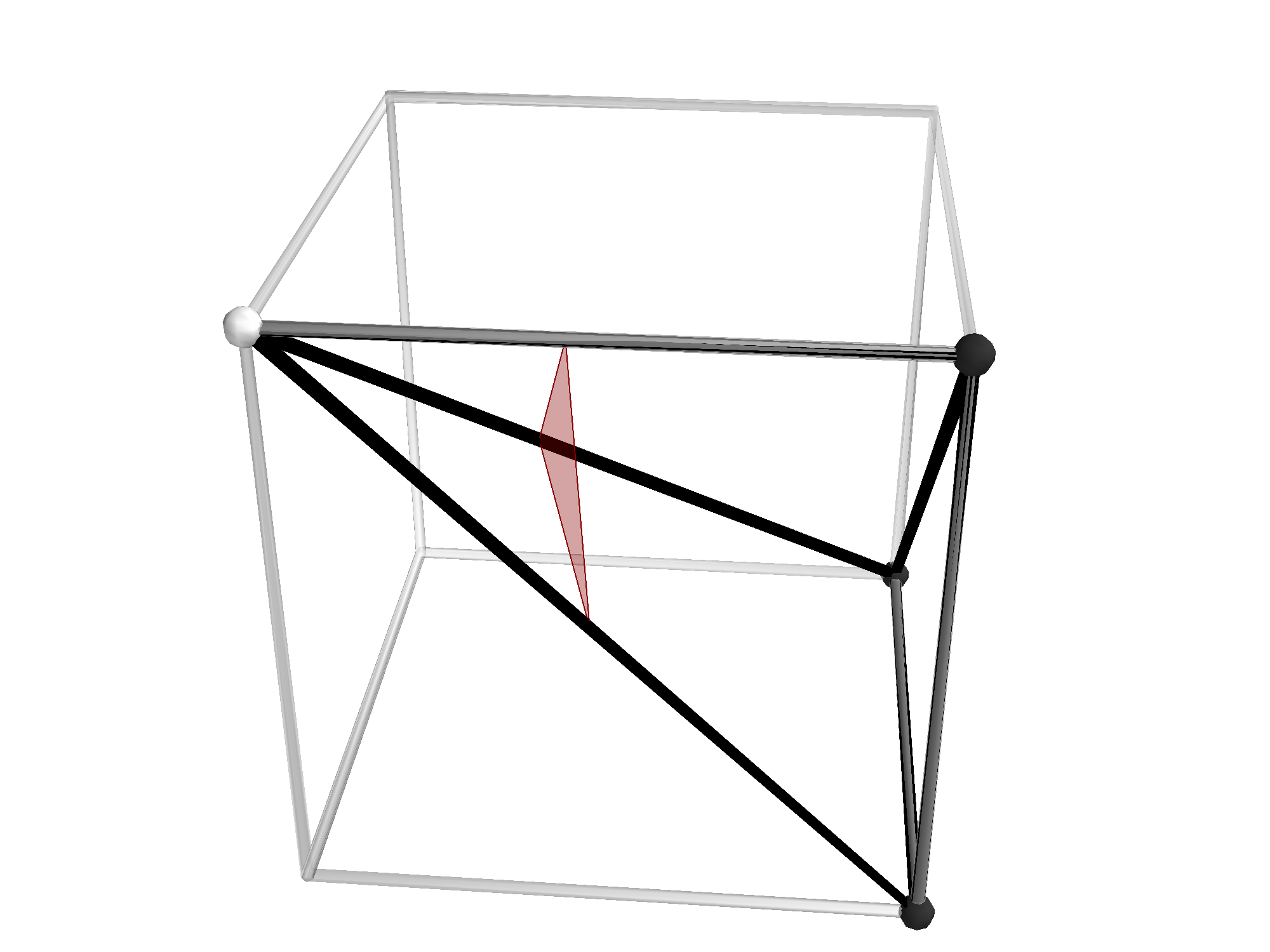}
  \includegraphics[width=0.32\textwidth]{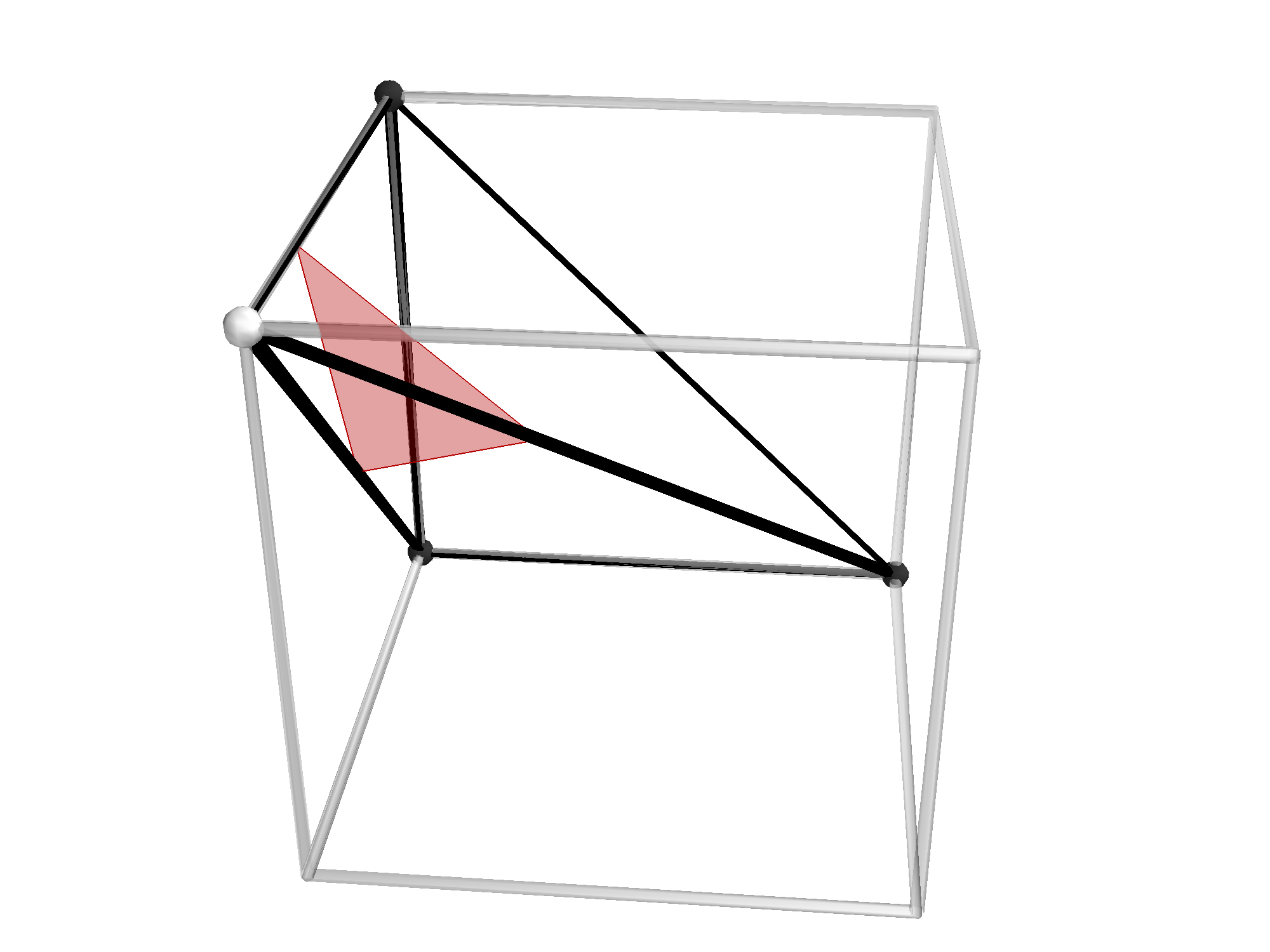}\\
  \caption{An example of our algorithm applied to a single pixel cube, displayed in the top left panel. The six smaller panels exhibit the six tetrahedra that we decompose the full cube into (shown as solid black lines). For each tetrahedron we generate the triangulated surface according to figure \ref{fig:2}. The final triangulated surface for this particular pixel is shown in the top right panel.}
  \label{fig:3}
\end{figure}

In figure \ref{fig:3} we exhibit an example of one complete pixel cube and the triangulation that our algorithm generates. The cube has five `in' and three `out' states, exhibited in the top left panel. The lower six panels exhibit the six tetrahedra that we generate as solid black lines, and the individual triangulations constructed within each. The resulting surface within the cube is exhibited in the top right panel. We repeat this construction for each pixel cube within the total volume, generating a closed triangulated surface encompassing the `in' states. 

The method is guaranteed to generate a closed surface. The topological ambiguities that are inherently present in the marching cubes algorithm \citep{Montani1994,marching_cubes_jgt} are eliminated by our choice of interpolation scheme, however as we discuss in appendix \ref{sec:error}, we introduce additional numerical uncertainty with our choice. The simplicity of the method relies on linear interpolation between tetrahedron vertices, however a more sophisticated algorithm should adopt bi-linear and tri-linear schemes when generating triangle vertices on the surface and interior region of a pixel cube respectively. There is no single unique bounding surface for a discrete density field, as the result will depend on the interpolation scheme adopted. We discuss the numerical error associated with our boundary reconstruction further in appendix \ref{sec:error}.

Once we have generated the triangulated surface, we can calculate the four scalar Minkowski functionals as 

\begin{eqnarray} \label{eq:w0} & & W_{0} = {1 \over V}\int_{Q} dV =  {1 \over V} \sum_{\cal T} \Delta V_{\cal T} \\ 
\label{eq:w1} & & W_{1} = {1 \over 6 V} \int_{\partial Q} dA = {1 \over 6 V} \sum_{\rm t} |t| \\
\label{eq:w2} & & W_{2} = {1 \over 3\pi V} \int_{\partial Q} G_{2} dA = {1 \over 6\pi V} \sum_{\rm e} |{\bf e}| \alpha_{\bf e} \\ 
\label{eq:w3} & & W_{3} = {1 \over 4\pi^{2} V} \int_{\partial Q} G_{3} dA = {1 \over 4\pi^{2} V} \sum_{\rm v} \left( 1 - {1 \over 2\pi} \sum_{{\rm T} \in {\bf v}} \phi_{\rm T}^{\bf v} \right)   \end{eqnarray}

\begin{figure}
  \centering
  \includegraphics[width=0.48\textwidth]{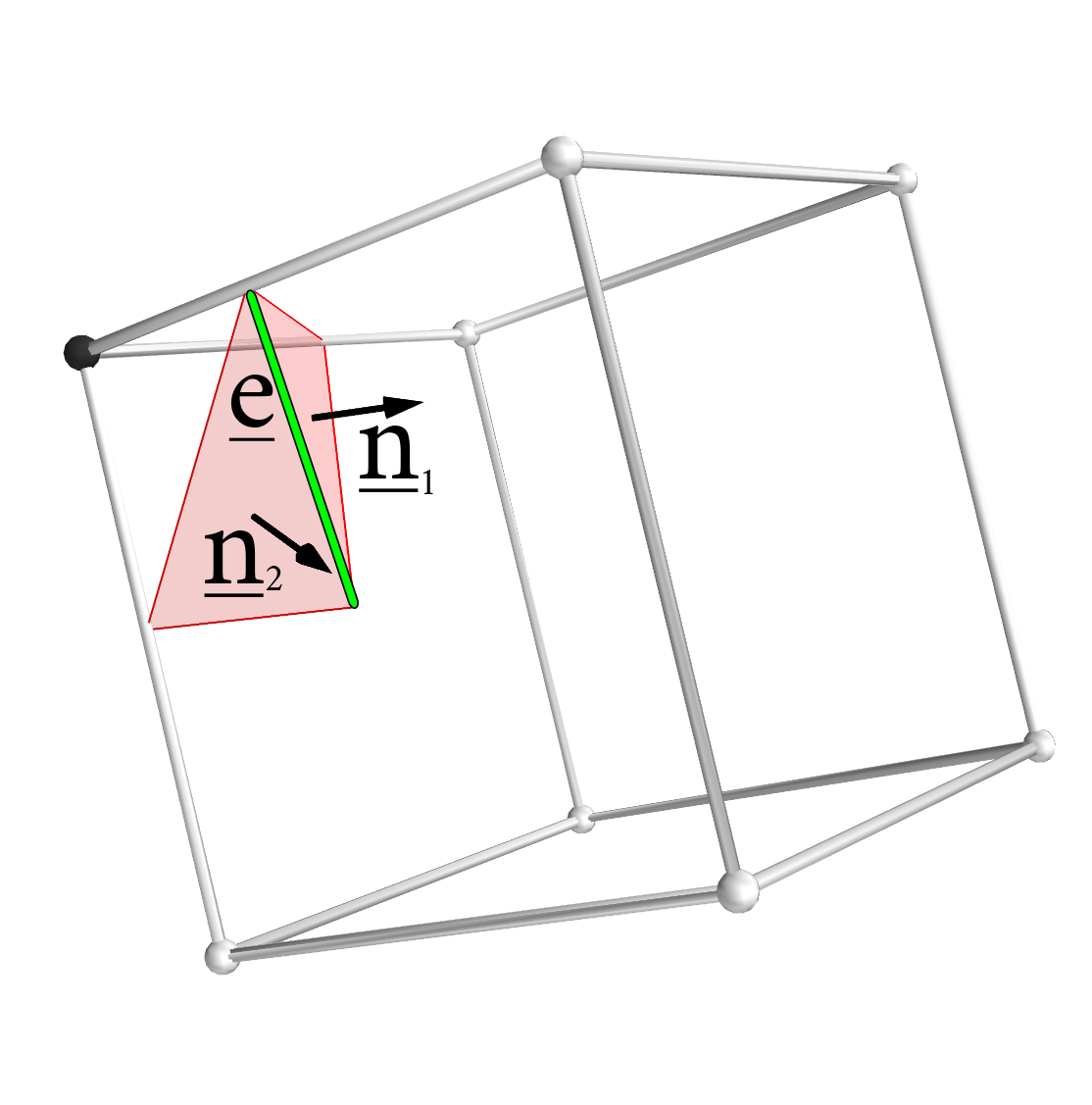}
  \includegraphics[width=0.48\textwidth]{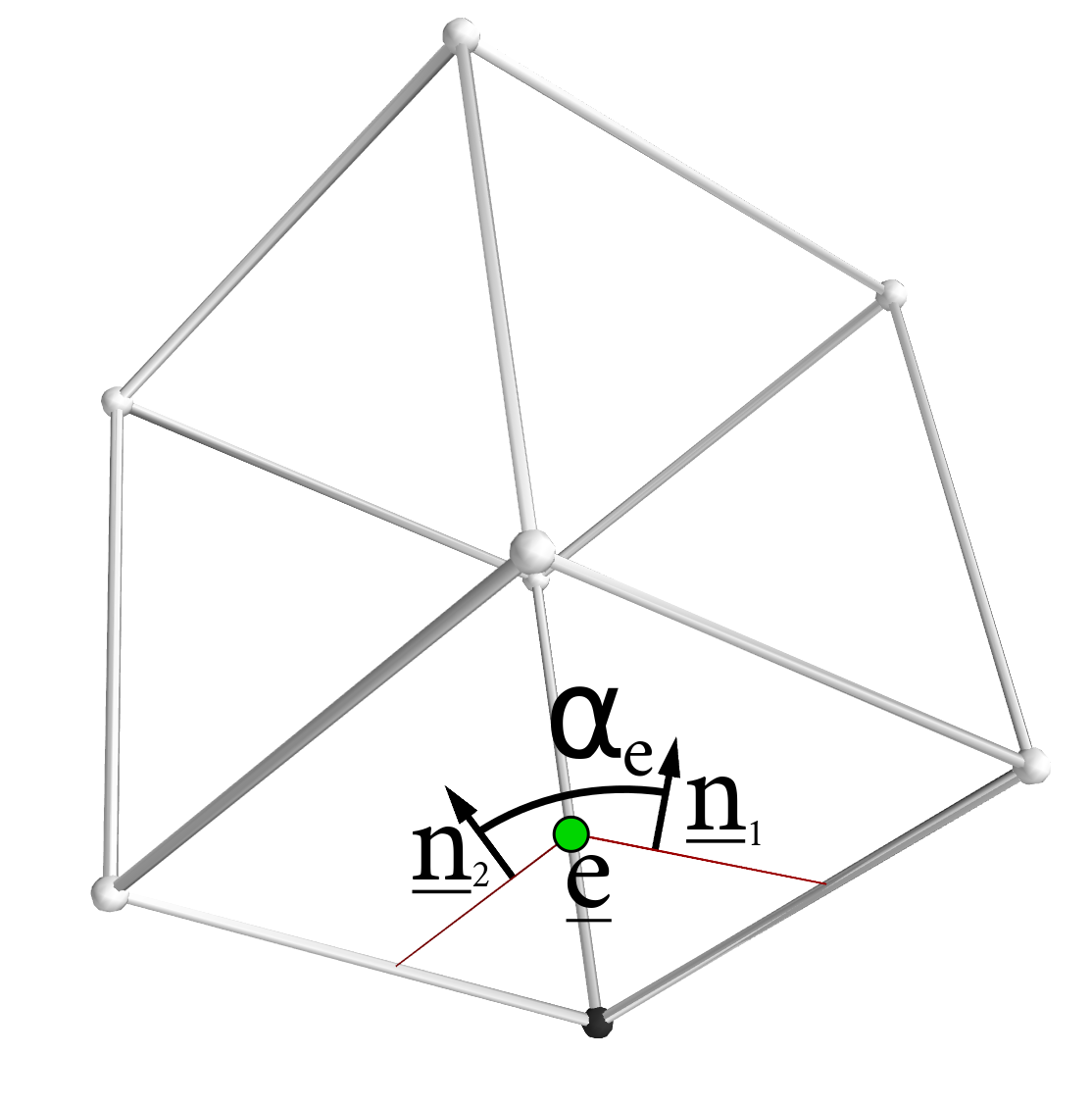}\\
  \caption{To generate the Minkowski functional $W_{2}$ and Minkowski tensor $W^{0,2}_{2}$ we reconstruct the angle $\alpha_{\rm e}$ from each edge in the triangulation. We exhibit $\alpha_{\rm e}$ for a single edge. In the left panel we display the surface constructed for a particular pixel cube. The edge ${\bf e}$, exhibited as a solid green line, joins two triangles with normals ${\bf n}_{1}$, ${\bf n}_{2}$. The quantity $\alpha_{\rm e}$ is the angle between ${\bf n}_{1}$, ${\bf n}_{2}$, in the plane perpendicular to the edge vector ${\bf e}$ (we exhibit this plane in the right panel, where we have rotated the cube in the left panel such that the line of sight is parallel to ${\bf e}$). }
  \label{fig:4}
\end{figure}

\noindent The discretized forms of $W_{0-3}$ - the expressions after the second equalities in equations ($\ref{eq:w0}-\ref{eq:w3}$) - represent sums over different quantities. $\sum_{\cal T}$ is a sum over each tetrahedron in our decomposition, and $\Delta V_{\cal T}$ is the volume occupied by the polygon defined by the triangle vertices and `in' states of each tetrahedron. That is, $\Delta V_{\cal T}$ is the volume enclosed by our triangulated surface within each tetrahedron. The sum $\sum_{\rm t}$ denotes the sum over all triangles within the bounding surface, where $|t|$ is the area of each triangle. $\sum_{\rm e}$ is the sum over all triangle edges ${\bf e}$, which have length $|{\bf e}|$. The angle $\alpha_{\bf e}$ represents the angle sub-tending the normals of two triangles that share edge ${\bf e}$ - an example is exhibited in figure \ref{fig:4}. 

\begin{figure}
 \centering
  \includegraphics[width=0.48\textwidth]{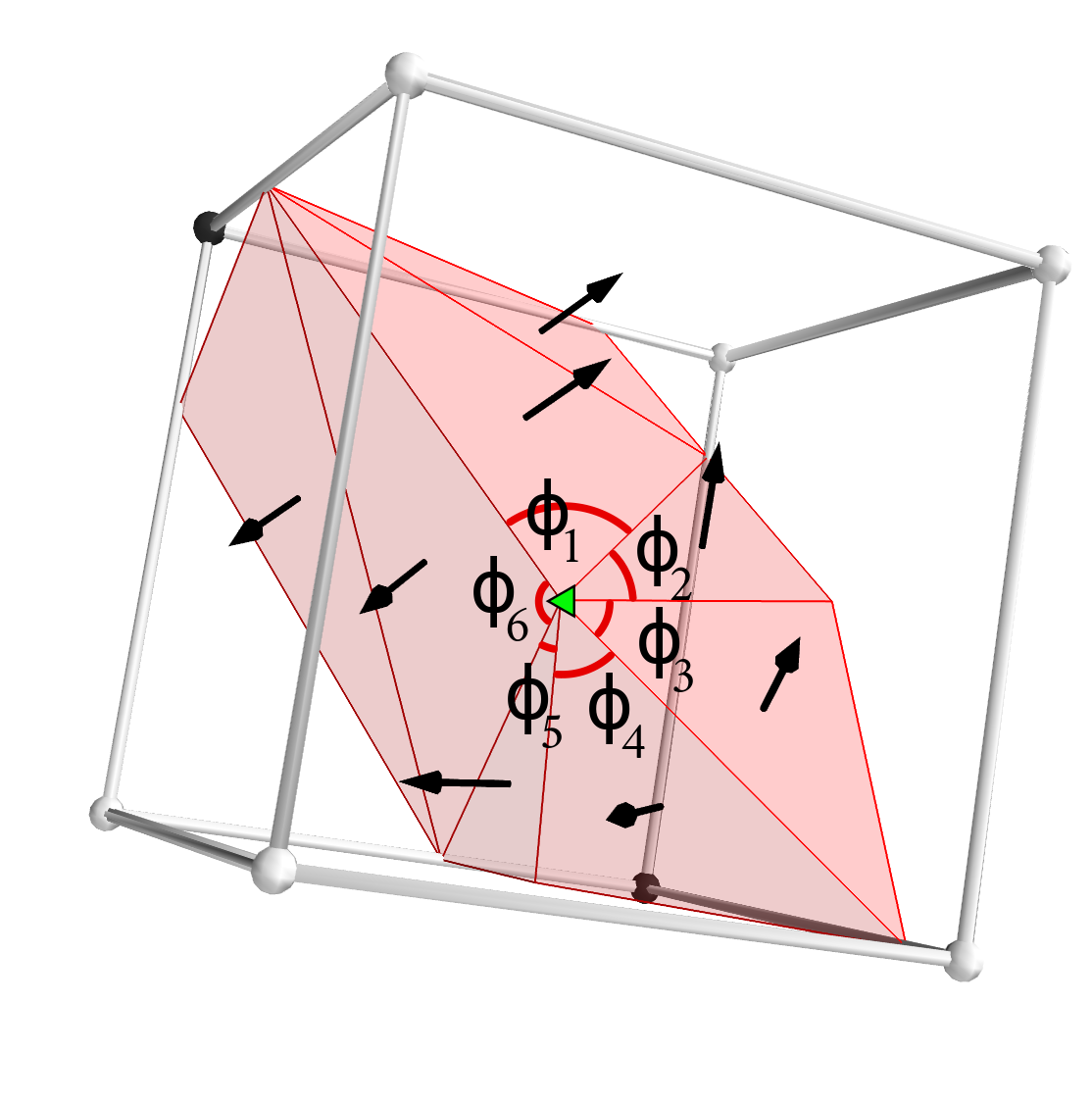}
  \caption{To generate the Minkowski functional $W_{3}$, we require the deficit angle at each vertex in the triangulated surface. An example is displayed - the green point in the center is a triangle vertex interior to this particular box. It is shared by six triangles in the surface - the deficit angle for this vertex is given by $2\pi - \sum_{i=1}^{6} \phi_{i}$.}
  \label{fig:5}
\end{figure}

The calculation of $W_{3}$ - the genus of the field - involves taking the sum of all interior angles of triangles that share a common vertex $v$, and subtracting $2\pi$ for each unique vertex in the triangulation. The calculation reduces to the sum of deficit angles at each vertex, as the Gaussian curvature at all other points on the triangulated surface (edges and triangle surfaces) is zero. An example is presented in figure \ref{fig:5} - we present a single cube containing a triangle vertex that lies within its interior (shown as a green triangle). The contribution of this vertex to the total genus of the excursion set is given by $2\pi - \sum_{i=1}^{6}\phi_{\rm i}$. We repeat this calculation for each triangle vertex in the bounding surface to generate $W_{3}$. The sum $\sum_{\rm v}$ in equation ($\ref{eq:w3}$) is the sum over all unique vertices in the boundary triangulation.

We can also calculate the Minkowski tensors from the bounding surface. They are given by \citep{1367-2630-15-8-083028}

\begin{eqnarray}\label{eq:mt021} & & (W^{0,2}_{1})_{ij} = {1 \over 6 V} \int_{\partial Q} {\bf \hat{n}}_{i} {\bf \hat{n}}_{j} dA = {1 \over 6 V} \sum_{t} |t| {\bf \hat{n}}_{i} {\bf \hat{n}}_{j} \\
\label{eq:mt022} & & (W^{0,2}_{2})_{ij} = {1 \over 3\pi V} \int_{\partial Q} G_{2} {\bf \hat{n}}_{i} {\bf \hat{n}}_{j} dA = {1 \over 6\pi V} \sum_{\rm e} |{\bf e}| \left( \left( \alpha_{\rm e} + \sin \alpha_{\rm e} \right) \left(\ddot{\bf n}_{\rm e}^{2}\right)_{ij} + \left( \alpha_{\rm e} - \sin \alpha_{\rm e} \right) \left(\dot{\bf n}_{\rm e}^{2}\right)_{ij} \right)  \end{eqnarray}

\noindent where we have introduced the additional vectors $\dot{\bf n}_{\rm e}$ and $\ddot{\bf n}_{\rm e}$, which are defined as $\ddot{\bf n}_{\rm e} = ({\bf \hat{n}} + {\bf \hat{n}}')/|{\bf \hat{n}} + {\bf \hat{n}}'|$ and $\dot{\bf n}_{\rm e} = \ddot{\bf n}_{\rm e} \times \hat{\bf e}$. $\hat{\bf e}$ is the unit vector pointing along an edge, and ${\bf \hat{n}}, {\bf \hat{n}'}$ are the unit normals of the two triangles that share the edge. We have written equations ($\ref{eq:mt021},\ref{eq:mt022}$) using index notation, where $_{i}$, $_{j}$ indices run over the standard Cartesian $x_{1},x_{2},x_{3}$ orthogonal coordinates.

\section{B. Sources of Numerical Error} 
 \label{sec:error}

We briefly review three sources of numerical error associated with our algorithm. The first is topological, the second morphological and the third regards the spurious anisotropy implicit within our interpolation scheme.

\begin{figure}
  \centering
  \includegraphics[width=0.45\textwidth]{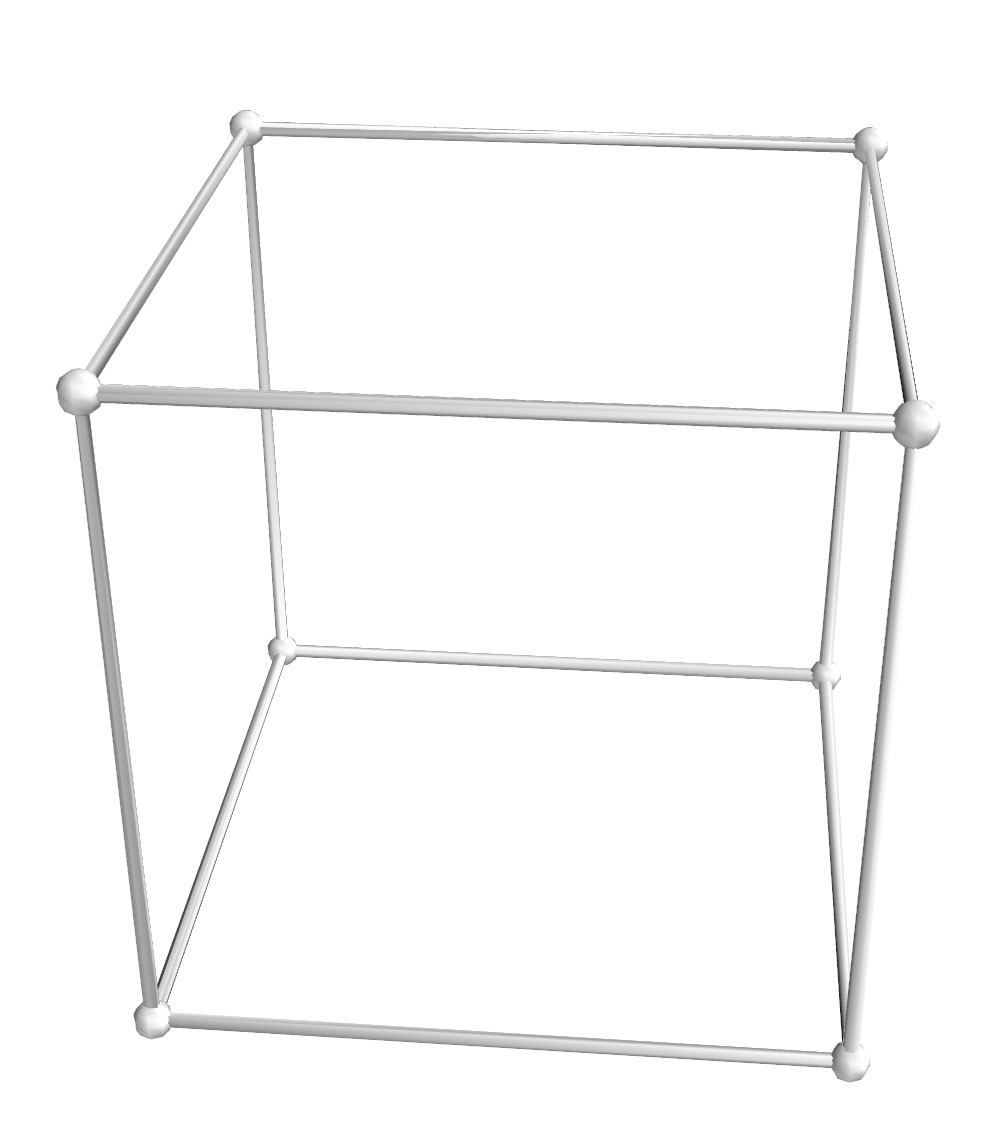}
  \includegraphics[width=0.45\textwidth]{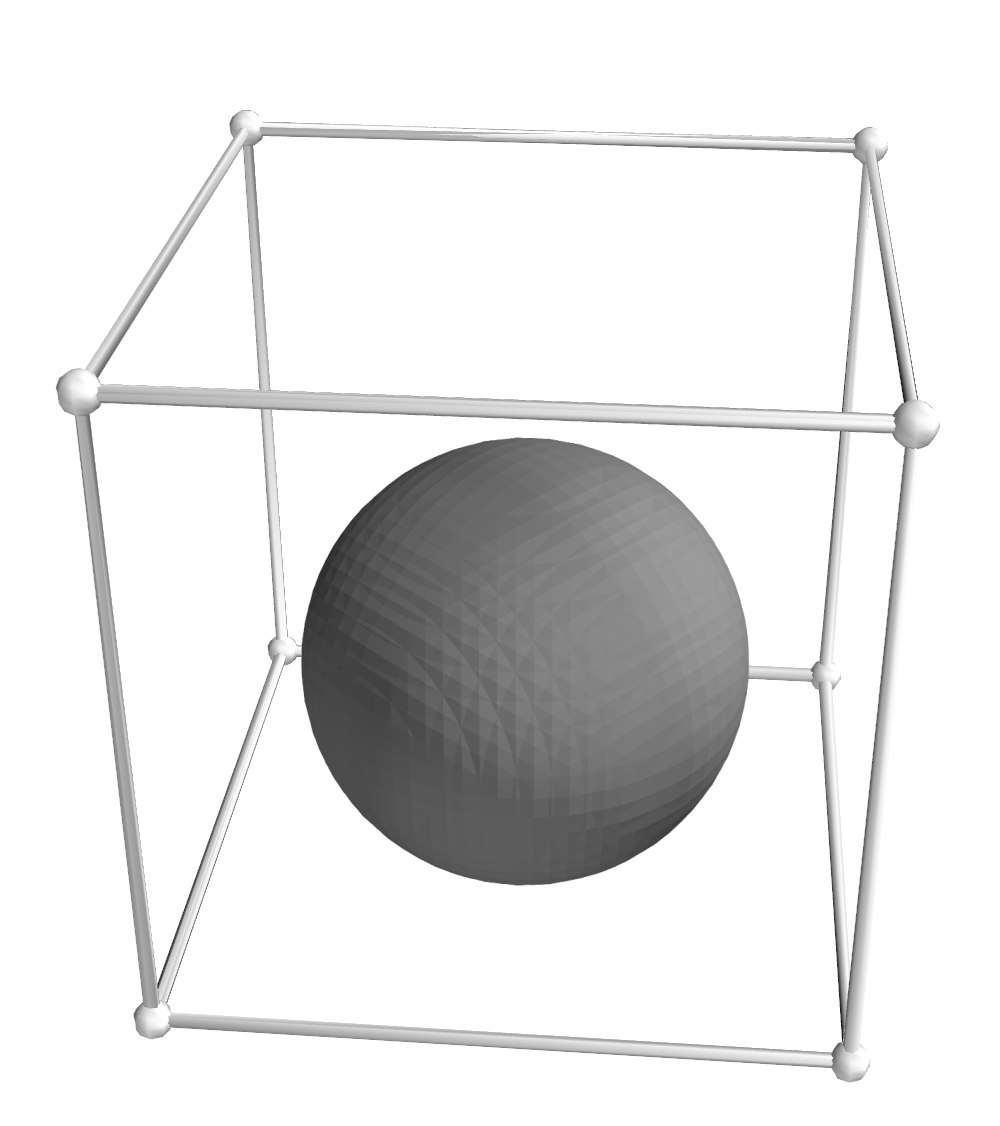}
  \caption{An example of the topological ambiguity that is present within our approach. We consider linear interpolation between vertices within a cube, and hence will miss small scale critical points which cannot be described with a linear scheme. In this example, the box has eight `out' vertices, and our algorithm will adopt the left panel in this case. However, a small scale peak (exhibited schematically in the right panel as a grey sphere) may be present, which would not be detected.}
  \label{fig:9}
\end{figure}

\subsection{B1. Topological Ambiguity}
\label{sec:topamb}

The topological inconsistency stems from the fact that the method adopted in this work involves linear interpolation of the density between points in a grid which assumes monotonicity of the field between grid points. However maxima, minima and saddle points are intrinsically higher order quantities. It follows that we will fail to detect structures that are present in the field that are of order of the pixel size. The cosmological density field will exhibit structure on all scales, and although small scale structures are suppressed by smoothing over several pixels they will still be present, particularly at high threshold values. An example is presented in figure \ref{fig:9} - applying our algorithm to a cube with eight `out' density pixels will always yield an empty box (left panel), but there may be small scale structure present (right panel). The solid grey sphere represents a peak in the density field, which is sub-resolution in size. The scenario exhibited schematically in figure \ref{fig:9} will modify the genus of the excursion set by one - at extreme threshold values $|\nu|$ these missing maxima and minima can comprise a significant fraction of the total. 

For a Gaussian field, one can provide an order of magnitude estimate of the number of density peaks that we will fail to detect using our approach. To do so we use the shape of a density field in the vicinity of a peak of height $\nu_{\rm pk}$, which is given in \citet{Bardeen:1985tr} as 

\begin{eqnarray} & & \nu \simeq \nu_{\rm pk} - {\sigma_{2} \over \sigma_{0}} {r^{2} \over 2} \left[ 1 + A(e,p)\right] x \\
& & A(e,p) = 3e \left[ 1 - \sin^{2}\theta \left(1 + \sin^{2}\phi \right) \right] + p \left(1 - 3 \sin^{2}\theta \cos^{2}\phi\right) \end{eqnarray}

\noindent where we have defined an arbitrary spherical coordinate system with $x_{3} = r \sin\theta \sin\phi$. The angle average of $A(e,p)$ is zero, and we take this limit - that is, we assume that peaks are spherically symmetric $\nu \simeq \nu_{\rm pk} - \sigma_{2} x r^{2} / (2\sigma_{0})$. Let us take $r$ as the radial distance between the location of a peak and the nearest grid point in a regular lattice at which we sample the field. The grid points are separated by distance $\epsilon$ and so $0 \le r \le \sqrt{3}\epsilon/2$, where $r=0$ is the case where the peak lies exactly on a sampled grid point and $r=\sqrt{3}\epsilon/2$ where the peak lies at the center of a pixel cube, maximally distant from any grid point. For our algorithm to fail to detect a peak, the field value $\nu$ must have dropped from $\nu_{\rm pk}$ at the location of the peak to below the threshold $\nu_{\rm t}$ at distance $r$ - that is $x$ must satisfy

\begin{equation} x > {2\sigma_{0} \over \sigma_{2}r^{2}} (\nu_{\rm pk}-\nu_{\rm t}) \end{equation}

\noindent The conditional probability that a point in the field takes value $x$ given that it is a peak of height $\nu_{\rm pk}$ is given by 

\begin{eqnarray} & & P(x|\nu_{\rm pk})dx = {e^{-(x-x_{*})^{2}/2(1-\gamma^{2})} \over \sqrt{2\pi (1-\gamma^{2})}} {f(x) dx \over G(\gamma,\gamma\nu_{\rm pk})} \\
& &  f(x) = {(x^{3} - 3x) \over 2} \left[ {\rm erf}\left[ \left({5 \over 2}\right)^{1/2}x\right] + {\rm erf}\left[ \left({5 \over 2}\right)^{1/2}{x \over 2}\right] \right] + \left({2 \over 5\pi}\right)^{1/2}\left[ \left( {31 x^{2} \over 4} + {8 \over 5}\right)e^{-5x^2/8} + \left( {x^{2} \over 2} - {8 \over 5}\right) e^{-5x^{2}/2}\right] \\
& & G(\gamma, x_{*}) =  \int_{0}^{\infty} dx f(x) {e^{-(x-x_{*})^{2}/2(1-\gamma^{2})} \over \sqrt{2\pi(1-\gamma^{2})}}  \end{eqnarray}

\noindent Therefore the number density of peaks that our algorithm will miss, as a function of threshold value $\nu_{\rm t}$ and distance $r$ from the nearest grid point, is given by 

\begin{equation} {\cal N}_{\rm missed}(r,\nu_{\rm t}) = \int_{\nu_{\rm t}}^{\infty} d\nu {\cal N}_{\rm pk}(\nu) \int_{x=2\sigma_{0}(\nu-\nu_{\rm t})/(\sigma_{2}r^{2})}^{\infty} P(x|\nu)dx  \end{equation}

\noindent In what follows we take the threshold range $0 < \nu_{\rm t} < 3$ and $0 < r \le \sqrt{3}\epsilon/2$. The quantity of interest is the total fraction of peaks that our algorithm will fail to detect as a function of threshold $\nu_{\rm t}$. Hence we construct the following statistic

\begin{equation} f_{\rm missed}(\nu_{\rm t}) = {\int_{0}^{\sqrt{3}\epsilon/2}dr {\cal W}(r) {\cal N}_{\rm missed}(r,\nu_{\rm t}) \over \int_{\nu_{\rm t}}^{\infty} {\cal N}_{\rm pk}(\nu)d\nu } \end{equation}

\noindent where ${\cal W}(r)$ is a weighting function that corresponds to the fractional volume within a pixel cube that is a distance $r$ from one of the vertices. ${\cal W}(r)$ is normalised as 

\begin{equation} \int_{0}^{\sqrt{3}\epsilon/2} {\cal W}(r) dr = 1 \end{equation}

\noindent We generate ${\cal W}(r)$ numerically, by decomposing a pixel cube into a regular $(200^3)$ lattice and calculating the number density of points that lie a distance $r$ from the nearest vertex. 

\begin{figure*}
  \centering
  \includegraphics[width=0.45\textwidth]{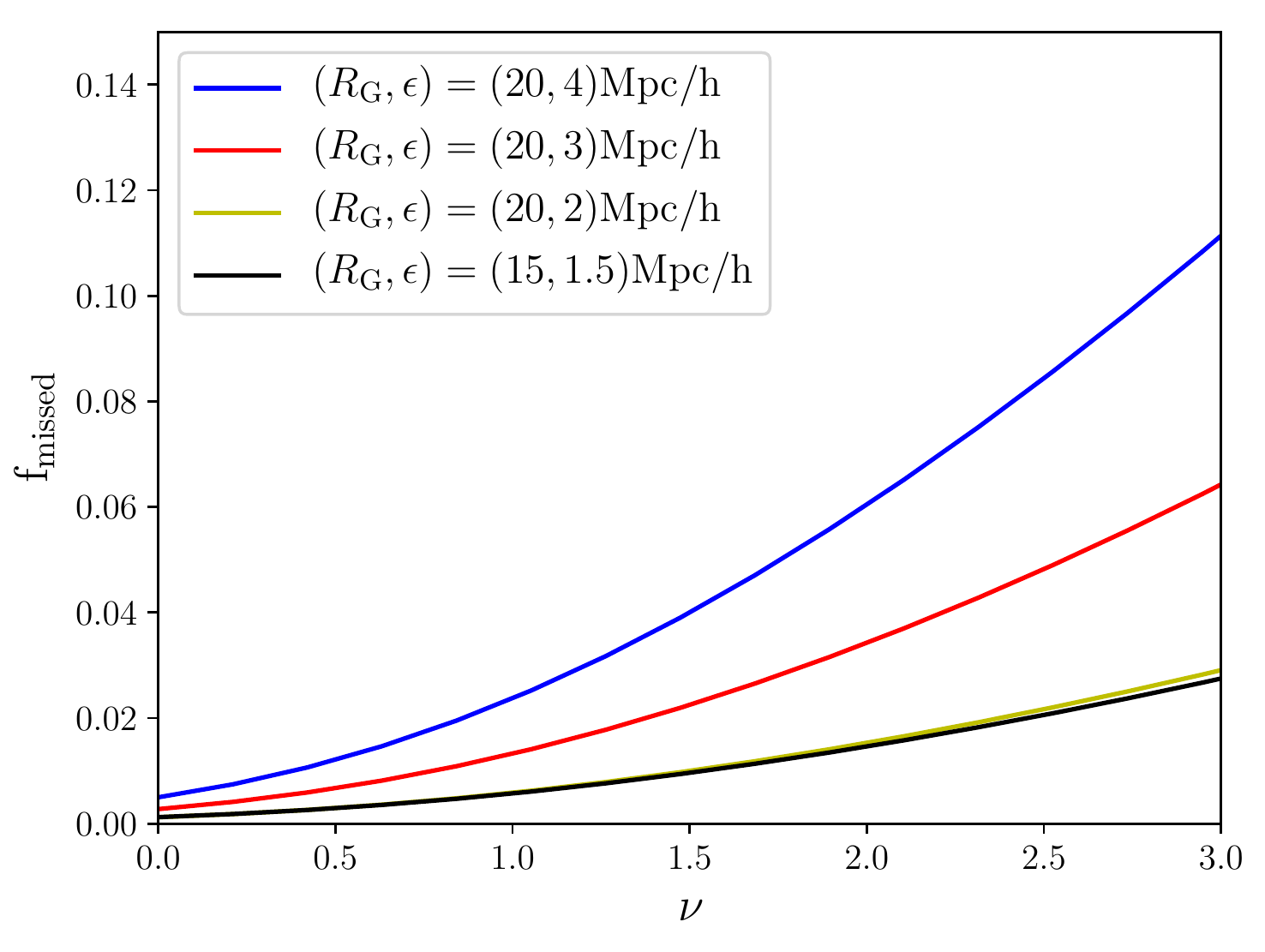} 
  \caption{The fraction of peaks that our algorithm will fail to detect as a function of threshold $\nu$, as discussed in appendix \ref{sec:topamb}. The blue, red, yellow and black lines correspond to smoothing scales and resolutions $(R_{\rm G},\epsilon) = (20,4), (20,3), (20,2), (15,1.5) h^{-1} \, {\rm Mpc}$ respectively. Smoothing over ten pixels will reduce the fraction of missed peaks to below $\sim 3\%$ for $|\nu| < 3$. }
  \label{fig:appb2}
\end{figure*}

We exhibit $f_{\rm missed}$ as a function of threshold $\nu$ for various $\epsilon$ values in figure \ref{fig:appb2}. We use a $\Lambda$CDM power spectrum to generate the field and vary $R_{\rm G}$ and resolution $\epsilon$. We find that the fraction of missed peaks increases with both $\nu$ and $\epsilon$ for fixed $R_{\rm G}$, as expected. One must smooth over ten pixels to ensure that the fraction of missed peaks remains below $f_{\rm missed} = 0.03$ for the threshold range $\nu < 3$ probed in this work. Although the missed peak fraction appears large for high threshold values, and this loss could potentially bias topological quantities, the statistical uncertainty on genus measurements increases with $|\nu|$. For example in this work we have used $V=(1024 h^{-1} \, {\rm Mpc})^{3}$ volumes and find a statistical error of $\Delta W_{3}/W_{3} \sim 7\%$ on the genus at $\nu=3$.

\subsection{B2. Discretization Error}\label{sec:dis_err}

\begin{figure}[!b]
  \centering
  \includegraphics[width=0.23\textwidth]{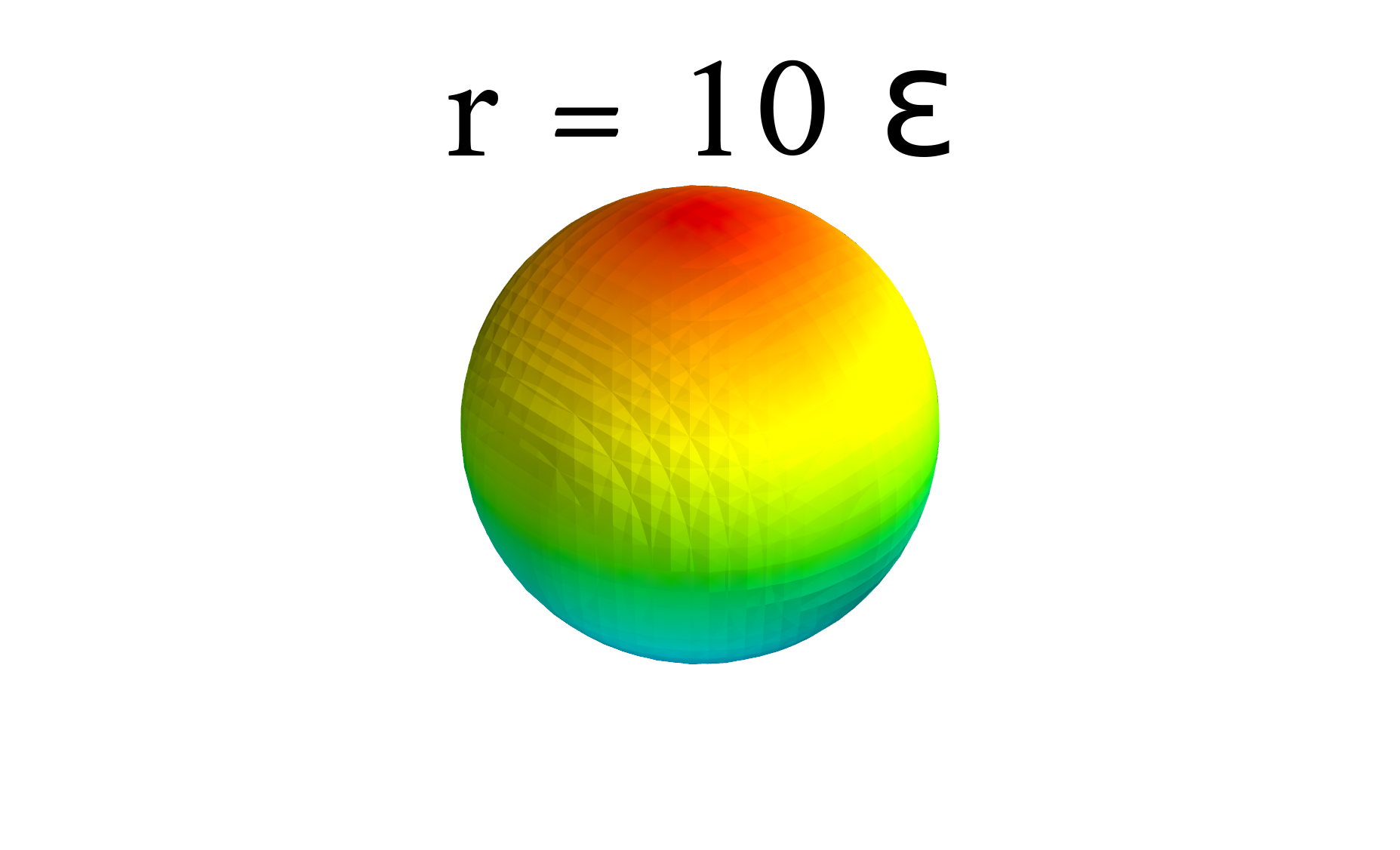}
  \includegraphics[width=0.23\textwidth]{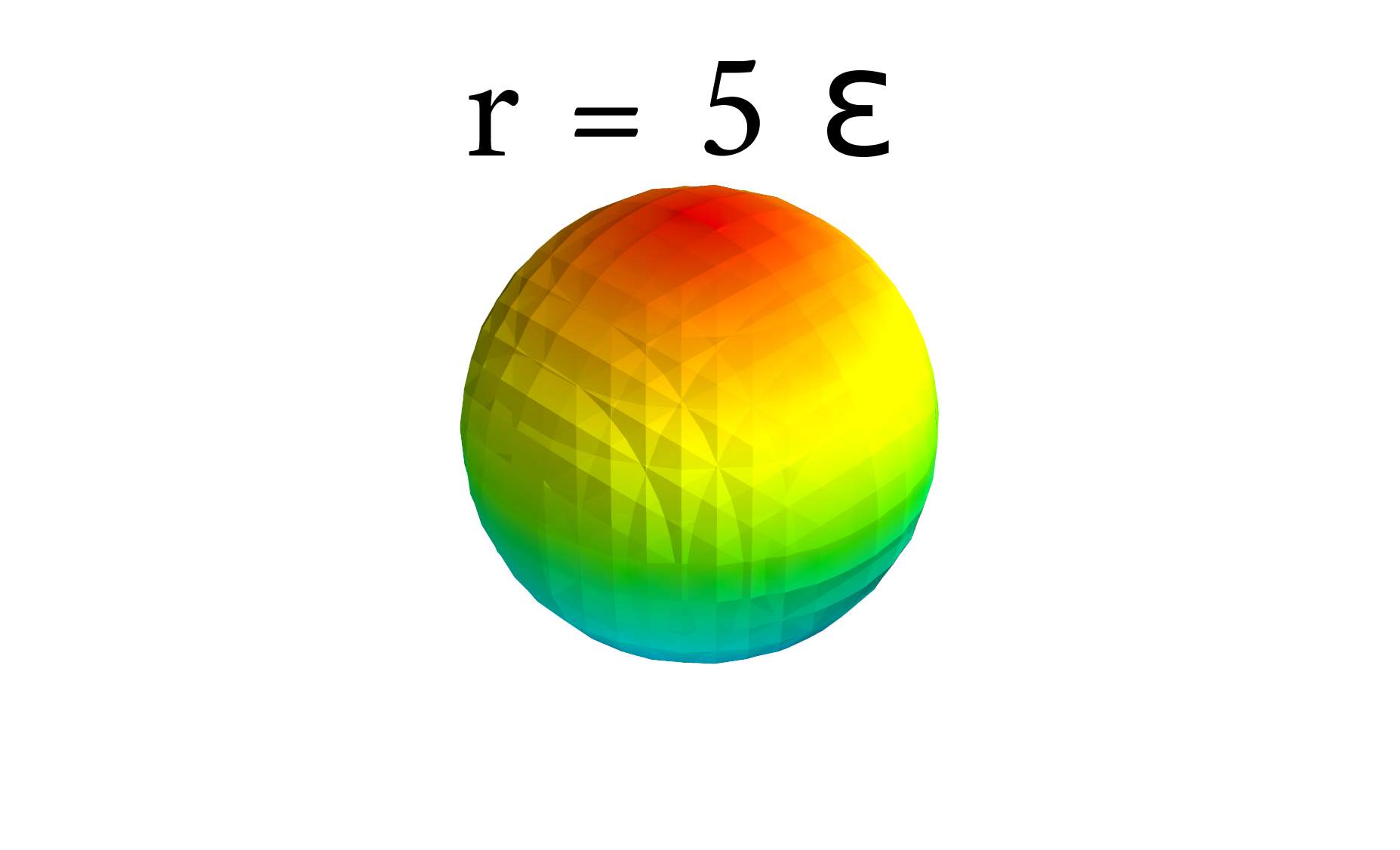}
  \includegraphics[width=0.23\textwidth]{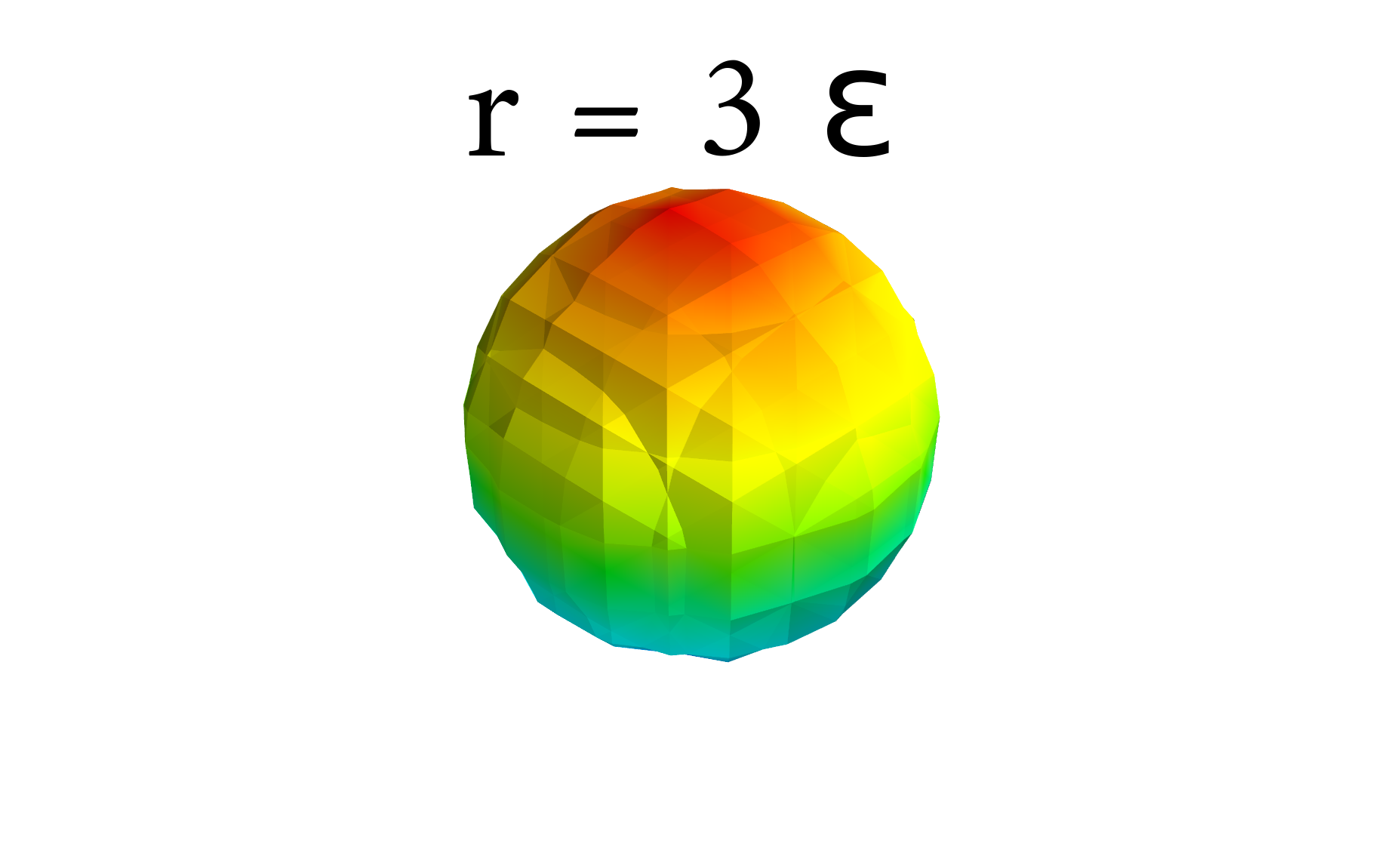}
  \includegraphics[width=0.23\textwidth]{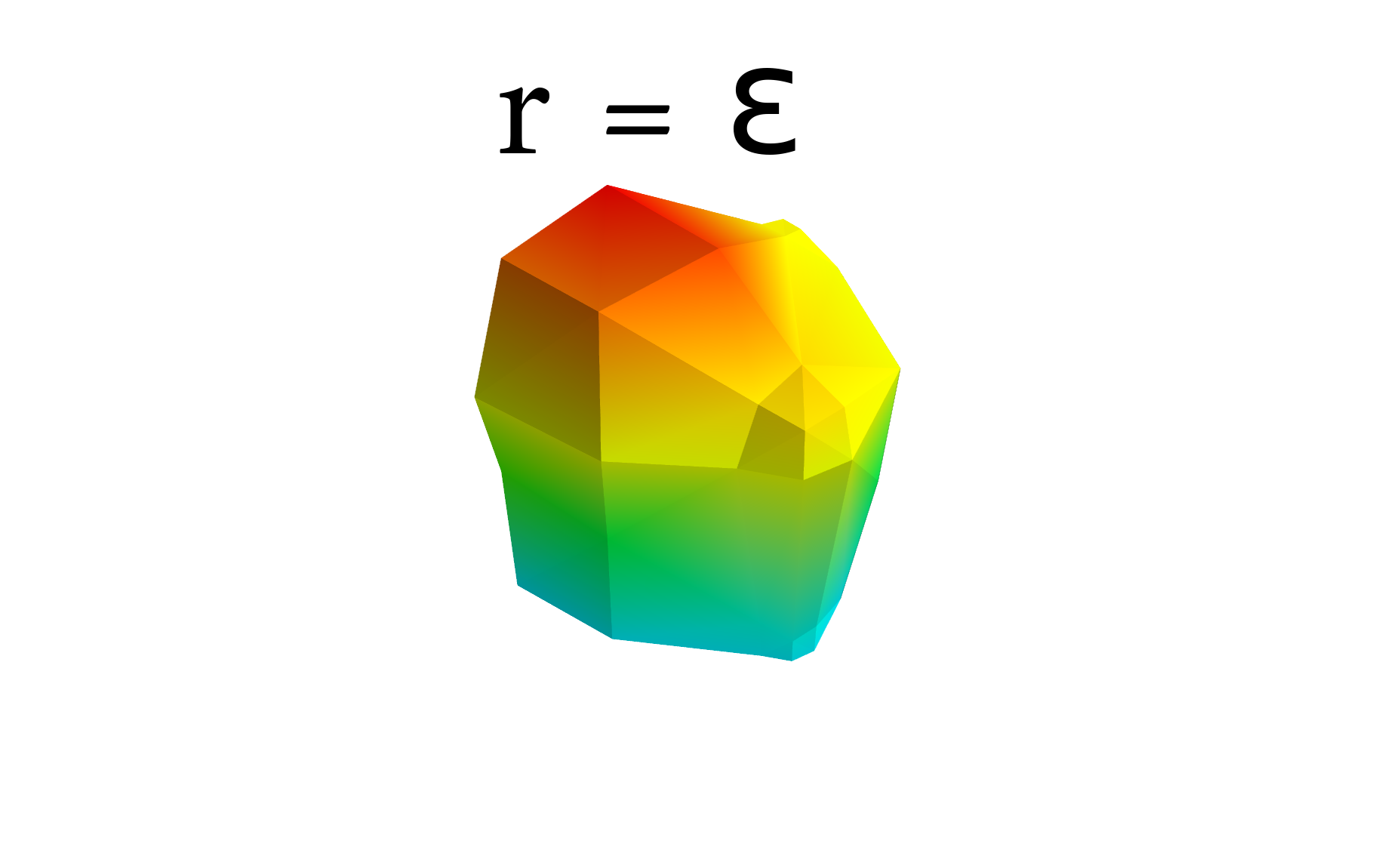}
  \caption{We exhibit the breakdown in accuracy of the marching tetrahedron algorithm. After generating a smooth spherical density field $\delta(x,y,z) = \delta(r)$ as in equation ($\ref{eq:df}$), we construct a surface of constant density around the central point. We decrease the radius of the sphere - from left to right, the radius is $r= 10, 5, 3, 1 \times \epsilon$, where $\epsilon$ is the resolution of the grid.}
  \label{fig:10}
\end{figure}

The second issue is the discretization of the bounding surface, which yields a polygonal structure that might not accurately represent the smooth underlying field. An example is exhibited in figure \ref{fig:10} - we have discretized a spherical density field of form 

\begin{equation}\label{eq:df} \delta(x_{1},x_{2},x_{3}) = {\delta_{\rm cen} \over  1 + |{\bf x} - {\bf x}_{\rm cen}|} \end{equation}

\noindent where ${\bf x} = (x_{1},x_{2},x_{3})$ and ${\bf x}_{\rm cen}$ is a random position vector that defines the centre of the density field. $\delta_{\rm cen}$ is the maximum value of the field, at ${\bf x} = {\bf x}_{\rm cen}$. We fix ${\bf x}_{\rm cen}$ and exhibit surfaces of constant threshold $\nu$ in figure \ref{fig:10}. As we increase the threshold $\nu$, we generate increasingly small spherical regions, and when the radius of the sphere approaches the resolution $\epsilon$ the triangulated mesh no longer accurately represents the underlying field. From left to right, the radius of the spheres in figure \ref{fig:10} are $r = |{\bf x} - {\bf x}_{\rm cen}| = 10 \epsilon, 5\epsilon, 3\epsilon, \epsilon$. One can observe the degradation in accuracy of the surface reconstruction with decreasing $r/\epsilon$.

\begin{figure}[!b]
  \centering
  \includegraphics[width=0.45\textwidth]{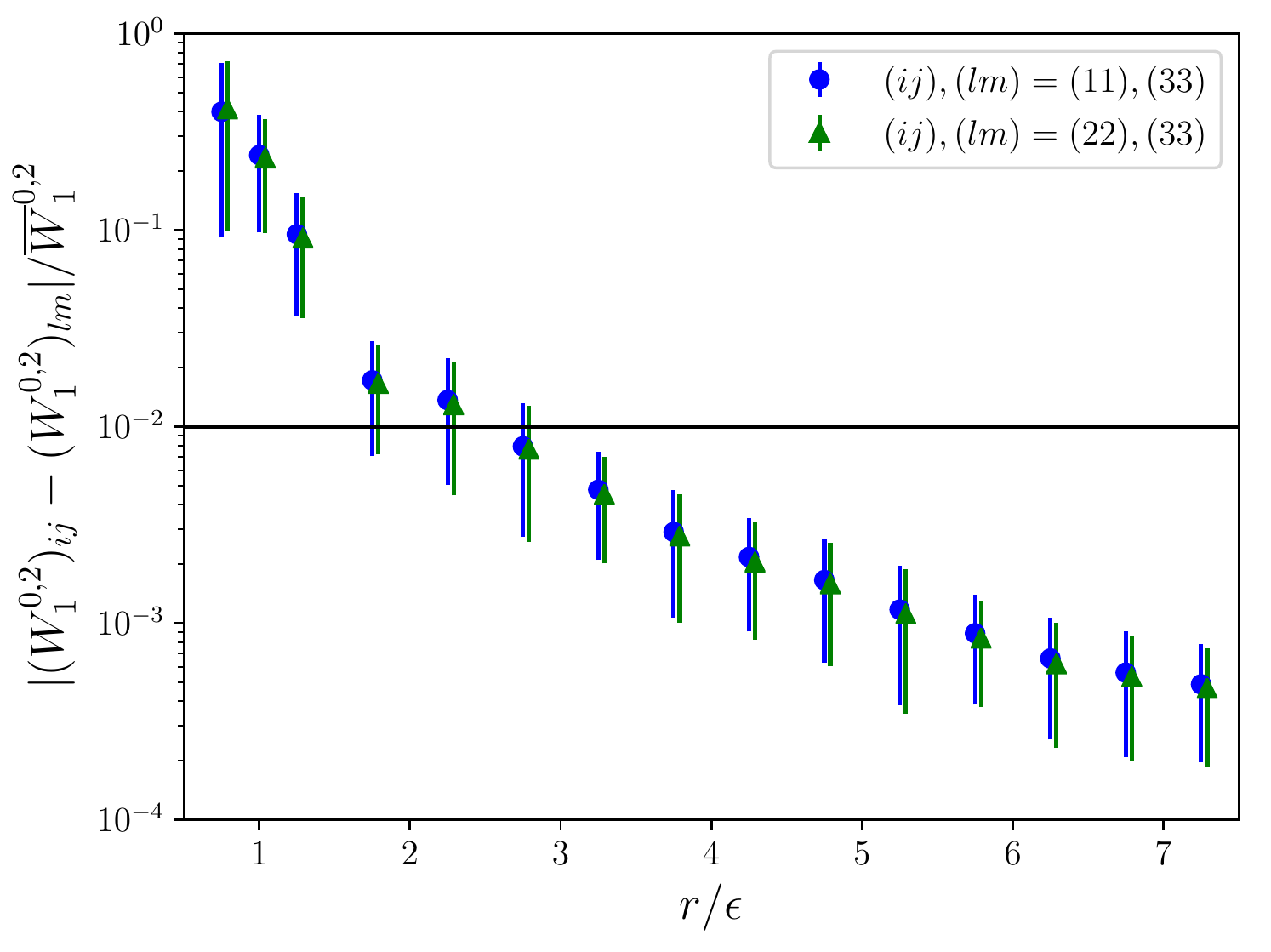}
  \includegraphics[width=0.45\textwidth]{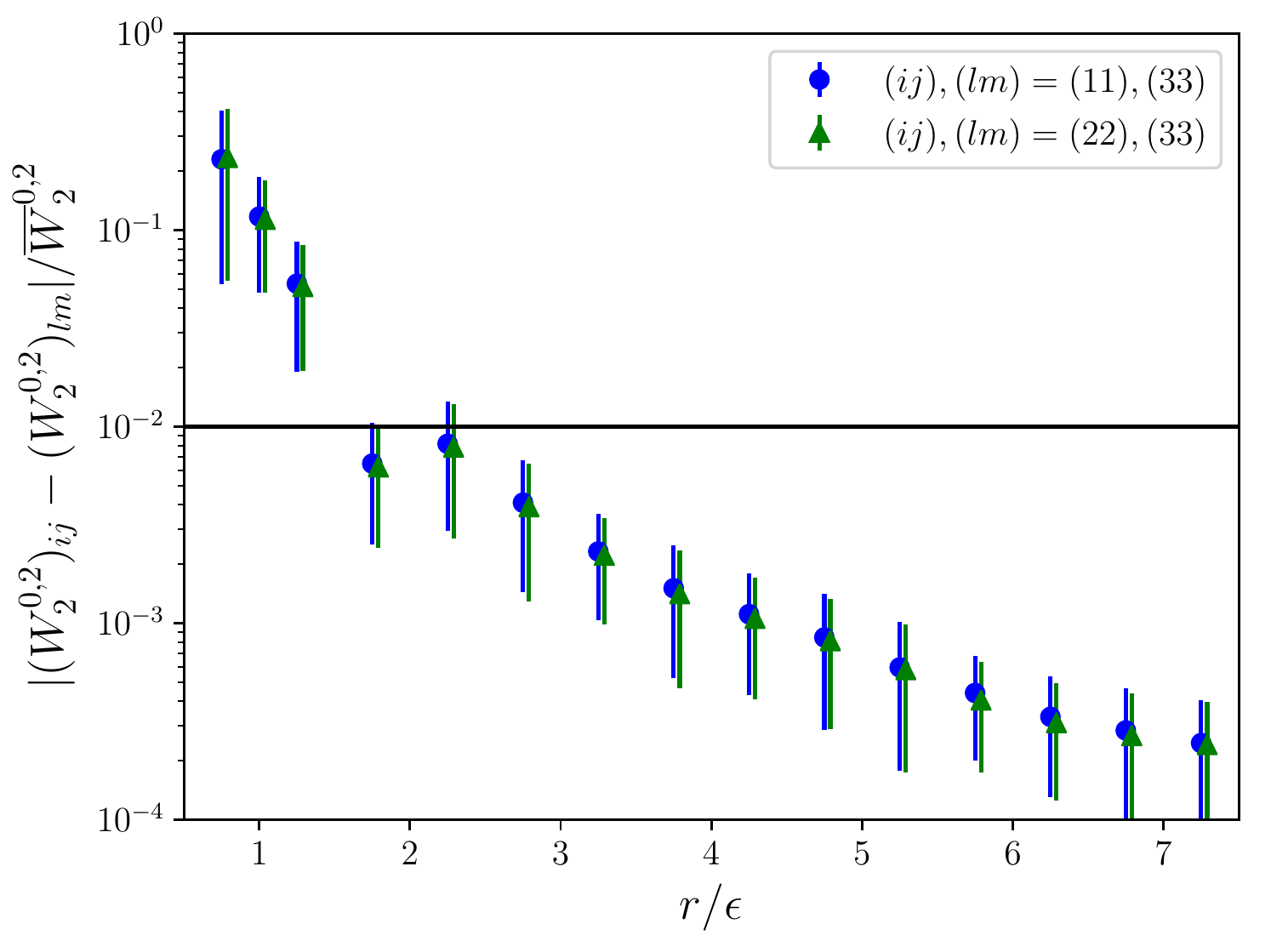}
  \caption{We calculate the Minkowski tensors $W^{0,2}_{1}$ (left panel) and $W^{0,2}_{2}$ (right panel) for $N_{\rm real}=50$ realisations of a randomly positioned spherical density field. We plot the absolute differences in diagonal components of the matrices divided by the mean value as a function of the radius of sphere in units of resolution $\epsilon$. All quantities should be consistent with zero for perfect spheres. For well resolved surfaces $r > 3\epsilon$, the statistics are consistent with their theoretical expectation to within $\sim 1\%$. However for poorly resolved objects the discrete nature of the bounding surface introduces spurious numerical anisotropy.}
  \label{fig:11}
\end{figure}

To quantify this effect, we generate $N_{\rm real} = 100$ density fields of the form ($\ref{eq:df}$), with fixed $\delta_{\rm cen}$ and random centres ${\bf x}_{\rm cen}$. We vary the threshold cut and consider how the properties of the spherical structure change as we decrease its resolution. Specifically we calculate the diagonal elements of the matrices $W^{0,2}_{1}$, $W^{0,2}_{2}$ for the spherical density field, which should satisfy $(W^{0,2}_{1})_{11} = (W^{0,2}_{1})_{22} = (W^{0,2}_{1})_{33}$ and $(W^{0,2}_{2})_{11} = (W^{0,2}_{2})_{22} = (W^{0,2}_{2})_{33}$ for an isotropic boundary. In figure \ref{fig:11} we exhibit the absolute difference between the diagonal matrix elements divided by the average value, as a function of the radius of the spherical overdensity normalised by the pixel resolution $r/\epsilon$. All quantities should be zero, and any departure is due to anisotropy generated from our discretization scheme. One can observe that the resolution effect is negligible for both statistics for well resolved objects $r > 3 \epsilon$, however objects that are of order of the pixel size can yield $\sim 20\%$ numerical error in our estimation of $W^{0,2}_{1}$, $W^{0,2}_{2}$. When calculating these quantities for stochastic fields, one must be careful to smooth over sufficiently large scales so that the excursion set is dominated by well resolved regions. 

\subsection{B3. Intrinsic Anisotropy}

The third numerical issue that should be addressed is the anisotropy associated with the interpolation scheme that we adopt. The method of marching tetrahedra decomposes the pixel cubes into six non-overlapping tetrahedra, and then performs linear interpolation between tetrahedron vertices that are `in' or `out' of the excursion set. This procedure breaks the symmetry of the cube along the major diagonal in which we decompose into tetrahedra. Poorly resolved excursion subsets will therefore exhibit a degree of alignment. The purpose of this work is to study anisotropic signals within the data - one must check that no spurious numerical artifacts are introduced. 

To test the reliability of the algorithm, we take the $N_{\rm real}=50$ realisations of a $\Lambda$CDM Gaussian random field used in section \ref{sec:4}, and calculate the Minkowski tensor $W^{0,2}_{1}$ for each connected region in the excursion set for each density field. We then calculate the corresponding eigenvectors and eigenvalues. For an isotropic field, the individual excursion subsets should exhibit no alignments, and the eigenvectors should be randomly directed on the unit sphere. Clustering in the eigenvectors will indicate artificial anisotropy generated by the method.

\begin{figure}[!b]
  \centering
  \includegraphics[width=0.45\textwidth]{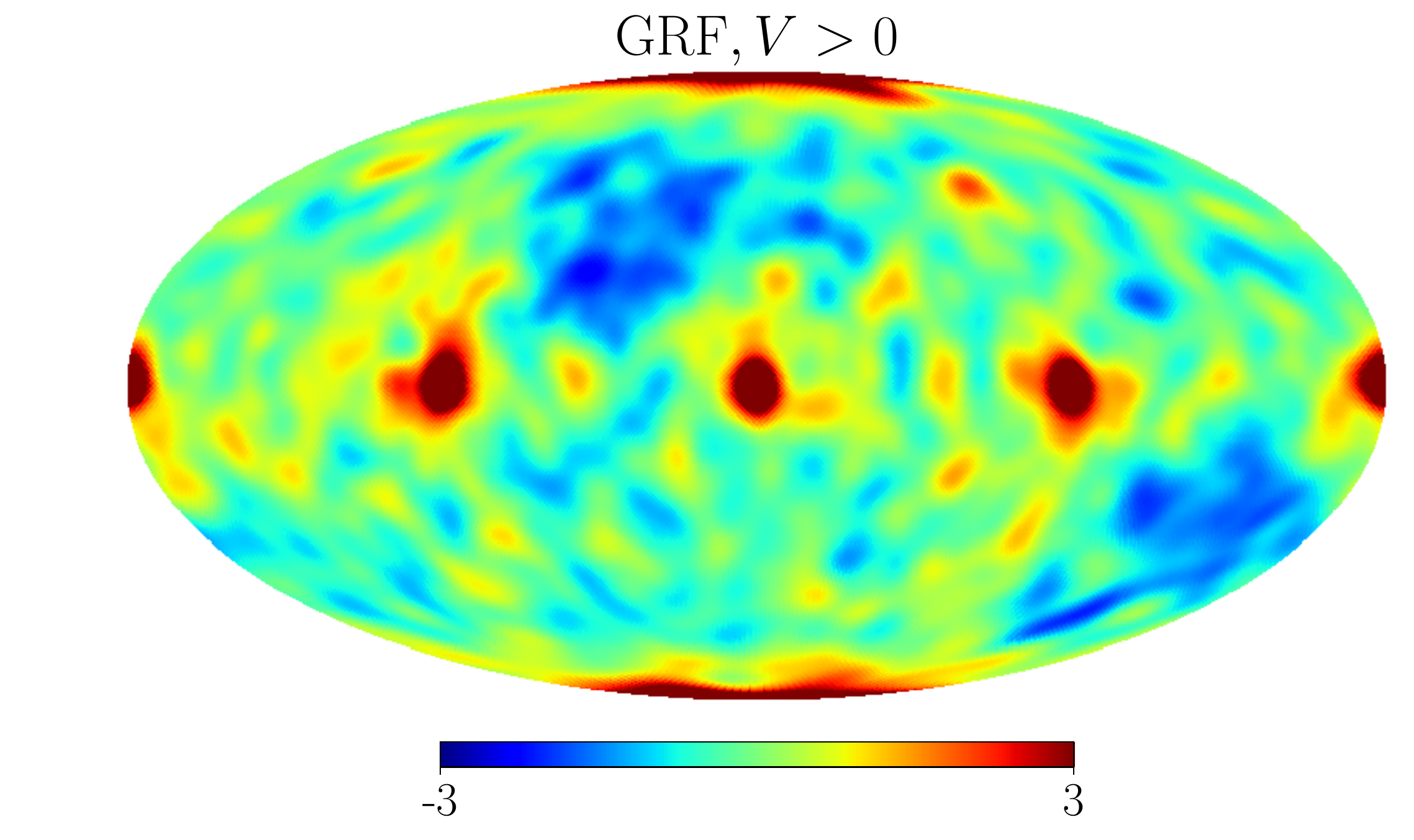}
  \includegraphics[width=0.45\textwidth]{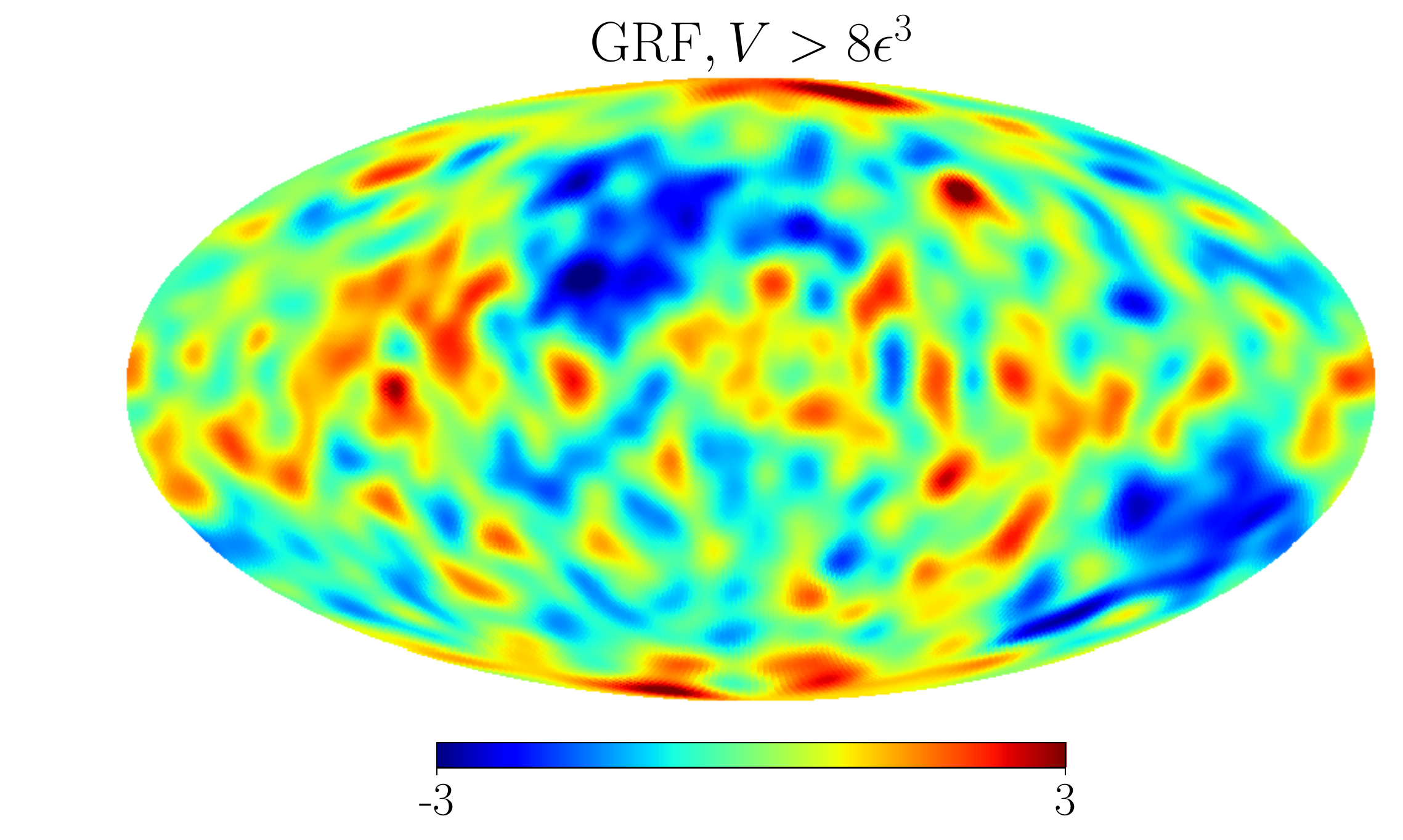}\\
  \includegraphics[width=0.45\textwidth]{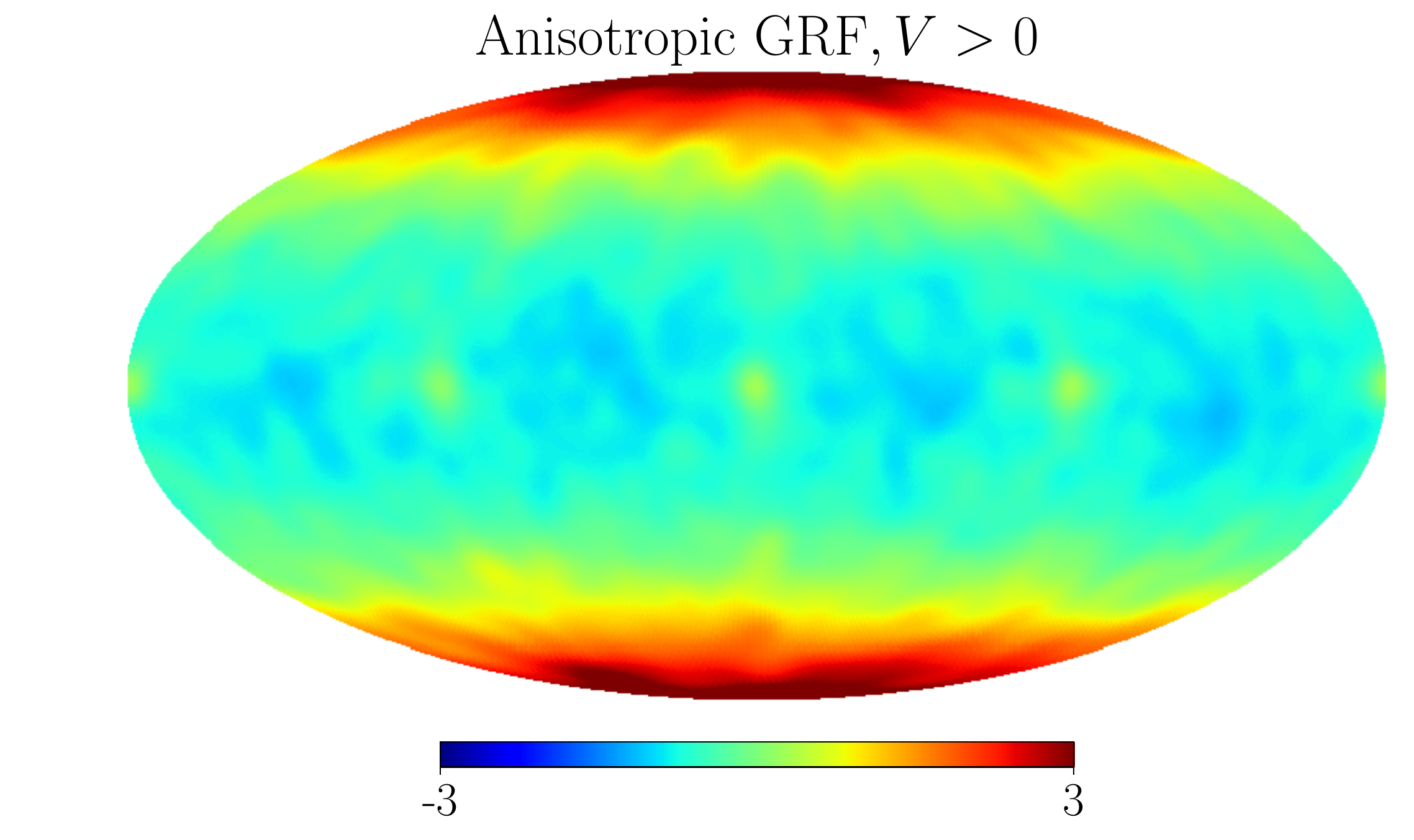}
  \includegraphics[width=0.45\textwidth]{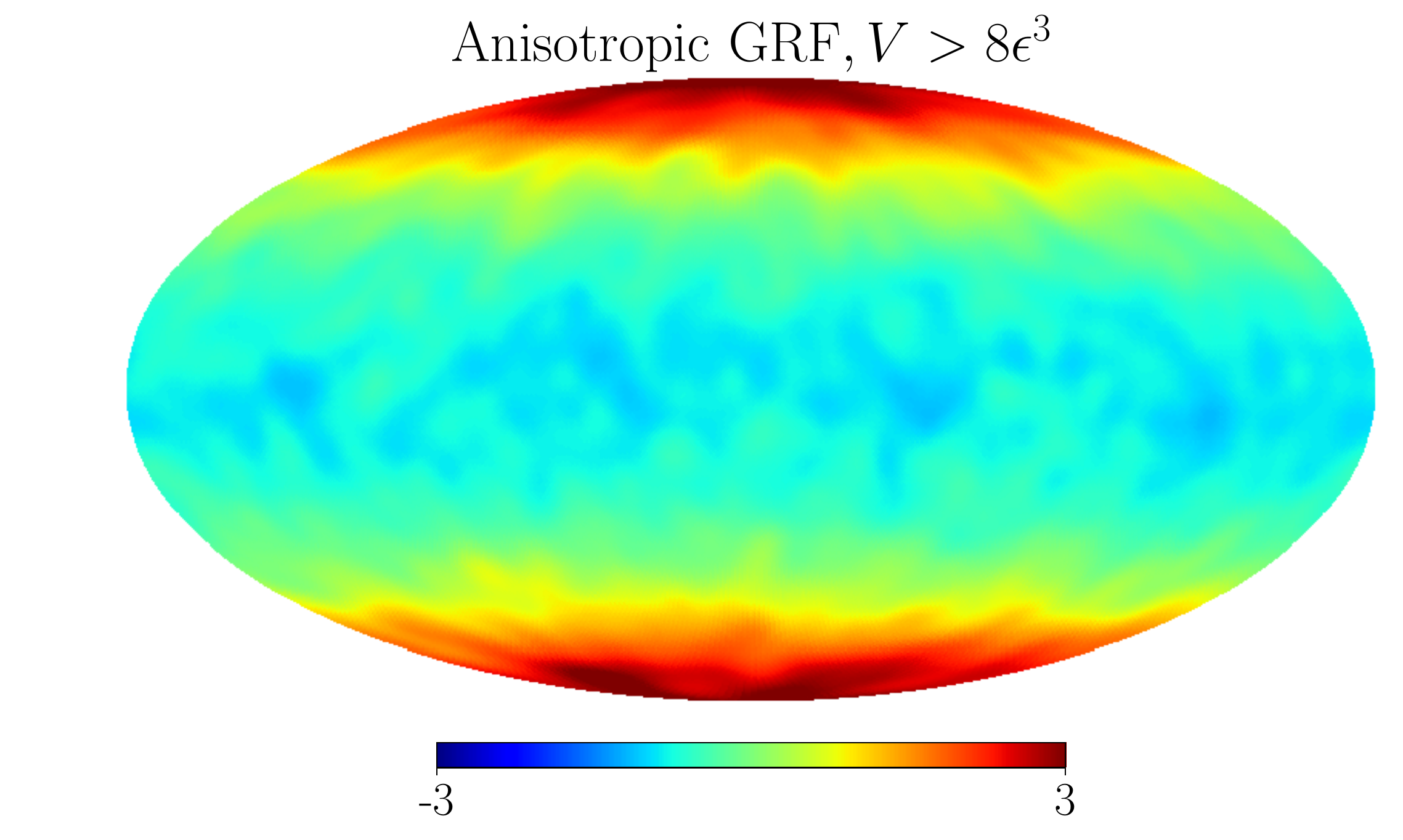}\\
  \caption{After generating $N_{\rm real} = 50$ realisations of a Gaussian random field, we collect all disjoint structures within the excursion sets and calculate their Minkowski tensors $W^{0,2}_{1}$. The eigenvectors of this matrix correspond to a set of axes along which the object is aligned. We exhibit the direction of the eigenvector corresponding to the largest eigenvalue for every distinct structure within the $N_{\rm real}=50$ Gaussian fields (top panels). Each individual structure will contribute a point on the sphere - we have smoothed this point distribution and generated Mollweide projections of the resulting distribution. For an isotropic Gaussian field, each eigenvector should be randomly directed. In the top left panel we exhibit all distinct subsets, and observe a distinct anisotropic pattern of six equi-spaced overdensities - two at the poles and four evenly spaced at the equator. This pattern indicates six directions along which poorly resolved subsets of the field will align. If we only consider objects that are well resolved, by cutting all structures with volume $V < 8 \epsilon^{3}$ (top right panel), then the spurious anisotropic signal is eliminated. For comparison, in the bottom panels we perform the same test on the anisotropic, Gaussian random field introduced in section \ref{sec:rsd}. In this case the physical anisotropic signal along the $x_{3}$ axis is clearly dominant regardless of the volume cut adopted. }
  \label{fig:moll}
\end{figure}

In figure \ref{fig:moll} we exhibit a Mollweide projection of the eigenvector corresponding to the largest eigenvalue of every connected region from the $N_{\rm real}=50$ realisations. Each connected region will generate an eigenvector direction as a point on the unit sphere - we create a smoothed distribution from the points, with hot spots indicating an overdensity of eigenvector pointings. The top left panel exhibits the distribution of all eigenvectors, and the top right panel only the eigenvector directions of objects with volume $V > 8 \epsilon^{3}$, where $\epsilon = 2 h^{-1} \, {\rm Mpc}$ is the resolution of our grid. That is, the top left panel exhibits all eigenvector pointings and the top right contains information on well resolved objects only. 

The top left panel exhibits clear anisotropy, with a set of six overdensities regularly spaced (two at the poles and four equi-spaced at the equator). It is clear that the unresolved objects generate an anisotropic signal that aligns with the Cartesian grid. However if we eliminate the unresolved regions from our average (top right panel), then we largely remove the spurious signal. 

We repeat our analysis using the $N_{\rm real}=50$ realisations of the anisotropic Gaussian field used in section \ref{sec:rsd}. In the bottom panels we exhibit the eigenvector pointings for the excursion subsets after applying no cut (left panel) and a volume cut $V > 8\epsilon^{3}$. The physical anisotropic signal in the $x_{3}$ direction is overwhelmingly dominant regardless of the cut. 

The spurious numerical anisotropic signal generated by the method is a form of noise that will provide a lower limit on the strength of any actual signal that can be detected using this algorithm. It is important to mitigate this numerical artifact by applying cuts to the data and removing poorly resolved regions. Pragmatically, one should study the robustness of the statistics as we change the total volume of the sample, the spatial resolution $\epsilon$ and the volume cut applied.

\section{C. Shape of Peaks in a Three Dimensional Gaussian Random Field}
\label{app:2}

The mean shape of a peak in a three dimensional Gaussian random field has been calculated in \citet{Bardeen:1985tr} - we briefly review this calculation, and relate the ellipticity and prolateness to the Minkowski tensors of individual excursion set sub-regions. 

For a Gaussian random field $F$, the probability distribution of $F$ and its first and second derivatives, $\eta_{i} = \nabla_{i} F/\sigma_{1}$ and $\nabla_{i}\nabla_{j}F$, respectively, at a point in three dimensional space, is given by a multivariate Gaussian probability distribution. We are interested in the shape of peaks of the density field, which are characterised by $\nabla_{i} F = 0$. If we diagonalise the second derivative matrix $\nabla_{i}\nabla_{j} F$ in terms of its eigenvalues $\lambda_{1,2,3}$, and then define $x= (\lambda_{1}+\lambda_{2}+\lambda_{3})/\sigma_{2}$, $y= (\lambda_{1}-\lambda_{3})/(2\sigma_{2})$, $z = (\lambda_{1} - 2\lambda_{2}+\lambda_{3})/(2\sigma_{2})$ and $\nu=F/\sigma_{0}$, we can define the joint probability as

\begin{eqnarray} & & P(\nu, {\bf \eta},x,y,z) d\nu d^{3}{\bf \eta} dx dy dz = {(15)^{5/2} \over 32\pi^{3}} {1 \over  (1-\gamma^{2})^{1/2}} |2y(y^{2}-z^{2})|e^{-Q} d\nu dx dy dz d^{3}\eta  \\ 
& & Q = {\nu^{2} \over 2} + {(x-x_{*})^{2} \over 2(1-\gamma^{2})} + {5 \over 2}(3y^{2}+z^{2}) + {3 \over 2} \eta . \eta  \end{eqnarray}

\noindent where $x_{*} = \gamma \nu$ and $\gamma = \sigma_{1}^{2}/(\sigma_{2}\sigma_{0})$. The cumulants of the field $\sigma_{i}$ are given by 

\begin{equation}  \sigma_{j}^{2} = \int {k^{2} dk \over 2\pi^{2}} P(k)k^{2j} \end{equation}

The conditional probability for the ellipticity $e = y/x$ and prolateness $p=z/x$ parameters, given a peak with parameters $\nu$ and $x$ is given by \citep{Bardeen:1985tr}

\begin{equation} P(e,p|\nu,x) de dp = {3^{2} 5^{5/2} \over (2\pi)^{1/2}} {x^{8}\over f(x)} e^{-5x^{2}(3e^{2} + p^{2})/2} W(e,p)de dp \end{equation}

\noindent where 

\begin{eqnarray} & &  W(e,p) = e (e^{2}-p^{2})(1-2p)\left[(1+p)^{2} - 9e^{2}\right] \chi(e,p) \\
\nonumber & & f(x) = {(x^{3} - 3x) \over 2} \left[ {\rm erf}\left[ \left({5 \over 2}\right)^{1/2}x\right] + {\rm erf}\left[ \left({5 \over 2}\right)^{1/2}{x \over 2}\right] \right] + \left({2 \over 5\pi}\right)^{1/2}\left[ \left( {31 x^{2} \over 4} + {8 \over 5}\right)e^{-5x^2/8} + \left( {x^{2} \over 2} - {8 \over 5}\right) e^{-5x^{2}/2}\right] \\
& &   \end{eqnarray}

\noindent and 

\begin{equation} \chi(e,p) =   
\begin{cases}
    1 & \text{if } 0 \le e \le 1/4 \text{ and } -e \le p \le e \\
    1 & \text{if } 1/4 \le e \le 1/2 \text{ and } -(1-3e) \le p \le e \\
    0 & \text{otherwise} 
\end{cases}
\end{equation}

\noindent the average peak density of maxima of parameters $\nu$, $x$ is given by 

\begin{equation} {\cal N}_{\rm pk}(\nu,x) dx d\nu = A e^{-\nu^{2}/2} f(x) e^{-(x-x_{*})^{2}/(2[1-\gamma^{2}])}  d\nu dx \end{equation}

\noindent where $A$ is an unimportant constant to which our final result will be independent. At each threshold level $\nu$, we estimate the mean shape of peaks that lie within the excursion set as 

\begin{equation}\label{eq:pep} P(e,p) = { \int_{\nu}^{\infty} \int_{0}^{\infty} P(e,p|\nu',x) {\cal N}_{\rm pk}(\nu',x) dx d\nu' \over \int_{\nu}^{\infty}\int_{0}^{\infty} {\cal N}_{\rm pk}(\nu',x) dx d\nu'} \end{equation} 

\noindent We perform the two dimensional integrals in ($\ref{eq:pep}$) and find the expectation values $e_{\rm m}$, $p_{\rm m}$ via $e_{\rm m} = \int \int \chi(e,p) P(e,p) e dp de$, $p_{\rm m} = \int \int \chi(e,p) P(e,p) p dp de$. 

We next relate the ellipticity and prolateness of a peak to the corresponding eigenvalues of the Minkowski tensor. For an ellipsoid, one cannot obtain a closed form expression for the eigenvalues of $W^{0,2}_{1}$ or $W^{0,2}_{2}$. One can write these quantities in terms of integrals over the parameterized ellipsoid surface as

\begin{eqnarray} \label{eq:intel1} & & \left( W^{0,2}_{1} \right)_{ij} = \int_{0}^{\pi} \sin u du \int_{0}^{2\pi} \sqrt{a^{2}b^{2}\cos^{2} u+c^{2}(b^{2}\cos^{2}v+a^{2} \sin^{2} v)\sin^{2} u} \hspace{2mm} \hat{n}_{i} \hat{n}_{j}  dv \\  
\label{eq:intel2} & & \left( W^{0,2}_{2} \right)_{ij} = \int_{0}^{\pi} \sin u du \int_{0}^{2\pi} \sqrt{a^{2}b^{2}\cos^{2} u+c^{2}(b^{2}\cos^{2}v+a^{2} \sin^{2} v)\sin^{2} u} \hspace{2mm} G_{2} \hat{n}_{i} \hat{n}_{j} dv  \end{eqnarray} 

\noindent where we have parameterized the surface of the ellipse as $(x,y,z) = (\sin u \cos v/a^{2}, \sin u \sin v/b^{2},\cos u /c^{2})$. In terms of this parameterization, the mean curvature $G_{2}$ is given by

\begin{equation}  G_{2} = {abc \left[ 3(a^{2} + b^{2})+2c^{2} + (a^{2}+b^{2}-2c^{2})\cos 2u - 2(a^{2}-b^{2})\cos 2v \sin^{2} u \right] \over 8 \left[a^{2} b^{2} \cos^{2} u +c^{2}(b^{2}\cos^{2} v +a^{2}\sin^{2} v ) \sin^{2} u \right]^{3/2}} \end{equation}

\noindent The most likely values of parameters $a,b,c$ are related to the ellipticity $e_{\rm m}$ and prolateness $p_{\rm m}$ as 

\begin{eqnarray}\label{eq:em} & & e_{\rm m} = {a_{\rm m}^{2} - c_{\rm m}^{2} \over 2(a_{\rm m}^{2} + b_{\rm m}^{2} + c_{\rm m}^{2})}   \\ 
\label{eq:pm} & & p_{\rm m} = { a_{\rm m}^{2} - 2b_{\rm m}^{2} + c_{\rm m}^{2} \over 2(a_{\rm m}^{2} + b_{\rm m}^{2} + c_{\rm m}^{2})}  \end{eqnarray}

\noindent therefore, for a given $e_{\rm m}, p_{\rm m}$ we invert the relations ($\ref{eq:em},\ref{eq:pm}$) and then calculate the integrals expressed in ($\ref{eq:intel1}, \ref{eq:intel2}$) for the corresponding $a_{\rm m},b_{\rm m}, c_{\rm m}$. There exists a redundancy in ($\ref{eq:em},\ref{eq:pm}$), so we eliminate $a_{\rm m}$ and define $\bar{b}_{\rm m} = b_{\rm m}/a_{\rm m}$, $\bar{c}_{\rm m} = c_{\rm m}/a_{\rm m}$. The ratio of eigenvalues calculated from the Minkowski tensors ($\ref{eq:intel1},\ref{eq:intel2}$) can be similarly defined purely in terms of $\bar{b}_{\rm m}$, $\bar{c}_{\rm m}$. We compare the theoretical values of $\beta^{(1)}_{i}$, $\beta^{(2)}_{i}$ obtained for peaks of the field with those of the excursion set regions, which are extended objects that may encompass multiple peaks. We expect agreement between theoretical expectation and numerical reconstruction in the high threshold limit $|\nu|$.

\bibliographystyle{ApJ}
\bibliography{biblio}{}

\end{document}